\documentstyle[12pt,aaspp4]{article}

\slugcomment{To Appear in Astrophysical Journal (20 August 1999)}

\begin {document}
\title{ L dwarfs and the substellar mass function}

\author {I. Neill Reid}
\affil {Palomar Observatory, 105-24, California Institute of Technology,
Pasadena, CA 91125;  e-mail: inr@astro.caltech.edu}

\author {J. Davy Kirkpatrick}
\affil {Infrared Processing and Analysis Center, 100-22, California Institute
of Technology, Pasadena, CA 91125}

\author {J. Liebert, A. Burrows}
\affil {Steward Observatory, University of Arizona, Tucson, AZ 85721}

\author {J. E. Gizis}
\affil {Department of Physics and
Astronomy, University of Massachusetts, Amherst, MA 01003}
 
\author {A. Burgasser}
\affil {Dept. of Physics, 103-33,  California Institute of Technology,
Pasadena, CA 91125}

\author {C. C. Dahn, D. Monet}
\affil {U.S. Naval Observatory, P.O. Box 1149, Flagstaff, AZ 86002}

\author {R. Cutri, C. A. Beichman}
\affil {Infrared Processing and Analysis Center, 100-22, California Institute
of Technology, Pasadena, CA 91125}

\author {M. Skrutskie}
\affil {Department of Physics and
Astronomy, University of Massachusetts, Amherst, MA 01003}

\begin{abstract}
Analysis of initial observations from sky surveys has shown that the resulting
photometric catalogues, combined with far-red optical data,
provide an extremely effective method of finding
isolated, very low-temperature objects in the general field. Follow-up
observations have already identified more than 25 sources with
temperatures cooler than the latest M dwarfs. A comparison with detailed 
model predictions (Burrows \& Sharp) indicates that these L dwarfs have 
effective temperatures between $\approx 2000\pm100$K and $1500\pm100$K, while
the available trigonometric parallax data place
their luminosities at between 10$^{-3.5}$ and 10$^{-4.3} L_\odot$. Those
properties, together with the detection of lithium in one-third of
the objects, are consistent with the majority having substellar masses. The mass function 
cannot be derived directly, since only near-infrared photometry and 
spectral types are available for most sources, but we can
incorporate VLM/brown dwarf models in simulations of the
Solar Neighbourhood population and constrain $\Psi(M)$ by comparing the
predicted L-dwarf surface densities and temperature distributions against
observations from the DENIS and 2MASS surveys.
The data, although sparse, can be represented by a power-law mass function,
$\Psi(M) \propto M^{- \alpha}$, with $1 < \alpha < 2$. Current results
favour a value nearer the lower limit. If $\alpha = 1.3$, then the 
local space density of $0.075 > {M \over M_\odot} > 0.01$ brown
dwarfs is 0.10  systems pc$^{-3}$. In that case brown dwarfs are twice as
common as main-sequence stars, but contribute no more than  $\sim15\%$ of the total
mass of the disk.
\end{abstract}


\section {Introduction }

The `stellar' mass function, $\Psi(M)$, describes the final product
of star formation, providing a global measurement of how diffuse gas is transformed
into quasi-static spheres in thermal and hydrostatic equilibrium, 
with masses from $\sim100M_\odot$ to $<0.1M_\odot$. 
The frequency of intermediate and high-mass stars produced by this
morphological restructuring determines galactic chemical 
evolution, while the proportion of material locked up in long-lived, low-luminosity 
dwarfs determines the overall mass-to-light ratio of a galactic population.
It is the latter characteristic which, with
the realisation of the importance of dark matter (Ostriker \& Peebles, 1973), focused
attention on Kumar's (1963) 'black dwarfs': objects whose central temperatures 
fail to cross the threshold for initiating hydrogen fusion. With
luminous lifetimes which are brief in astronomical timescales, such objects might constitute
substantial, but invisible, repositories of baryonic matter.

Relabelled as brown dwarfs (Tarter, 1976), these `failed stars', 
have been the target of numerous surveys over the last two decades. Those projects
reached fruition in only the last few years, with the identification of 
substellar mass objects in the Pleiades (Rebolo et al, 1995), as stellar companions
(Nakajima et al, 1995) and in the general field (Ruiz et al, 1997; Tinney, 1998).
Until recently, however, statistical analyses have been hampered by the relatively
small numbers of detected objects, particularly in the field where surveys were 
forced to rely on wide-field imaging at optical wavelengths. That circumstance
has changed with the inception of the near-infrared DENIS (Epchtein et al, 1994)
and 2MASS (Skrutskie et al, 1997) sky surveys.

VLM stars have effective temperatures of less than 3000 K, while brown dwarfs
spend most of their life at temperatures below 2000 K. As a result, the peak of
the energy distribution lies at wavelengths in the 1-3$\mu m$ region, where the only available
large-scale survey is the pioneering Two Micron Sky Survey (Neugebauer \& Leighton,
1969), barely reaching 3rd magnitude in the K band. DENIS and 2MASS extend
the sensitivity limits at near-infrared wavelengths
by factors of more than 10$^4$ to K$_S \sim$13.5 and K$_S \sim$15 magnitude
respectively (where K$_S$ is the K-short passband defined by Persson et al, 1998).
Both projects have succeeded in identifying extremely red, ultracool dwarfs 
 (Delfosse et al, 1997; Kirkpatrick et al, 1999a - hereinafter K2ML) 
with optical spectra similar to the previously-known VLM dwarfs GD 165B (Kirkpatrick
et al, 1999b) and Kelu 1 (Ruiz et al, 1997). Those sources clearly mark an extension
of the M-dwarf sequence to lower temperatures, and K2ML have codified that
progression in defining the new spectral class, type L. 

The present paper uses new data  provided by these surveys to probe the
form of the mass function at and below the hydrogen-burning limit, comparing
the DENIS and 2MASS observations against model predictions. 
Our analysis models the substellar mass function as a continuous  extension of
the stellar mass function, $\Psi_*(M): M > 0.08 M_\odot$. A prerequisite
for those calculations, therefore, is an accurate description of $\Psi_*(M)$,
serving as an anchor for our simulations. With that in mind, section 2 updates
the census of stars and stellar systems within 8 parsecs of the Sun and defines
$\Psi_*(M)$ in the range 0.1 to 1.0 M$_\odot$. Section 3 describes how
theoretical predictions of (L, T$_{eff}$) as a function of mass and age
are matched against the (K, spectral type) data for local L dwarfs;
section 4 outlines the structure of the simulations; section 5 summarises
the results; and section 6 presents our conclusions.

\section {The stellar mass function}

\subsection {The nearby star census}

Until recently, most studies of the nearest stars followed van de Kamp (1972)
in limiting the 'Solar Neighbourhood' to a 5.2-parsec (17 light year) radius
sphere - a volume which encompasses only 45 stellar systems. The extensive 
surveys undertaken over the last three decades, notably spectroscopy of
proper-motion stars and intensive radial velocity and high-resolution imaging
searches for binary companions, allow present-day analyses to consider
a larger sample - although still restricted to declinations accessible from
northern observatories (i.e. by Luyten at Palomar and Henry \& McCarthy at Steward). Reid \& Gizis
(1997 - RG97) laid the foundations of this reference sample, collating literature
data for candidate nearby stars, particularly spectroscopy by Kirkpatrick et al (1995)
and Reid et al (1996) of late-type stars from the (never published) preliminary version of the
Third Nearby Star Catalogue compiled by H. Jahreiss and W. Gliese.

The RG97 nearby-star compilation 
lists 106 stellar systems, including 151 stars, which are both north
of $\delta = -30^o$ and, at the time of publication, had assigned distance of less than 
8 parsecs from the Sun. The latter distances 
were based on a weighted combination of trigonometric parallax measurements
and spectroscopic parallaxes. Completeness is clearly of paramount importance
in these calculations, and RG97 present an extensive discussion of the available
observations. They conclude that 
the sample is likely to be statistically complete at the $\sim90\%$ level. In particular,
they note that while an extrapolation of the 5.2-parsec luminosity function predicts
$\sim30$ additional $r < 8$ pc mid-type M-dwarfs, all would be expected to have m$_r < 16$, 
well above the magnitude limit of available proper-motion catalogues. Nonetheless, since
this marks the first attempt to extend coverage to these larger distances, one expects
modifications to the sample as new, more accurate data are obtained. The Hipparcos
astrometric catalogue (ESA, 1997) provides such observations for many systems in the
initial 8-parsec sample, while more detailed searches for binary (and
tertiary) companions have also been
undertaken. We have used these new 
results to produce a revised northern 8-parsec catalogue as our baseline sample.

Hipparcos measurements eliminate four single stars and two binaries from the RG97
8-parsec sample. Two other binaries, Gl 185 and Gl 831, are formally outwith the 
8-parsec limit if one adopts the Hipparcos parallaxes outright, but lie just within the
limit in a weighted combination of ground-based and Hipparcos data.
Of the 108 stars in 100 systems with Hipparcos parallaxes
$\pi_H > 0".125$, eighteen are south
of $\delta = -30^o$  and five are northern stars with parallax measurements
of low precision\footnote{ The five stars in question are BD-13:637B, Gl563.2A
and B, BD+24:3192B and BD-15:6346B. Spectroscopy with the Double Spectrograph
on the Hale 200-inch (Reid, in prep.) identifies all five as K-type stars {--}
clearly incompatible with the M$_V > 12.5$ inferred from $\pi_H$.}.
Those stars are not included in the present analysis. However, the 
improved Hipparcos  astrometry adds several confirmed M-dwarfs to the 8-parsec sample, 
each with parallax measured to milliarcsecond precision. Table 1 summarises
the resultant additions and subtractions.

Newly-discovered binary and tertiary companions further modify the sample.
Delfosse et al (1999a) have identified additional components in three of
the systems listed in RG97 (Table 1), while Henry et al (1997, 1999) and
Oppenheimer et al (1999) use higher-resolution imaging to identify
new members in five systems (two in common with Delfosse et al).
These new discoveries eliminate two stars from the sample:  
LP 476-407 was resolved as visual binary by Henry et al's (1997) speckle imaging, 
and the primary component is a double-lined spectroscopic binary (Delfosse et al, 1999a), 
while infrared speckle measurements show that G89-32 is an equal-luminosity 
binary, separation 0.7 arcseconds (Henry et al, 1997).
In both cases the fainter apparent magnitude of each component leads to 
a smaller inferred spectroscopic parallax, removing the systems beyond the
8-parsec limit.

The other four systems with newly-discovered companions
remain within the 8-parsec sample
since all have trigonometric parallax measurements.
Gl831 and Gl 896 were identified previously as binaries, while 
RG97 list Gl 829 as a single star. Henry et al and Delfosse et al have discovered one
new companion in each, reclassifying the systems as
two triples and a binary respectively\footnote{
Delfosse et al also show that the previously-known companion in the SB1 binary G203-47 
is a white dwarf, giving a total of nine degenerates in the
sample - four single stars and five companions.}. Finally, Oppenheimer 
et al's observations have revealed a third component in the 
LTT 1445 (LP 771-95/96) system. 

Including both the brown dwarf Gl 229B and the Sun, the revised 
northern 8-parsec  catalogue includes 150 objects in 103 systems: 68
single stars, 25 binaries, 9 triples and one quadruple\footnote{RG 97 considered
Gl 643/Gl 644 as a quintuple, but Hipparcos astrometry show that Gl 643 is nearer by 0.75
parsecs. Given this substantial separation, we group the five stars as a
single and a quadruple system.}. All save four systems have distances 
derived from accurate trigonometric
parallax measurements, 68 based on Hipparcos data. 
The overall multiple star fraction (i.e. the fraction of
stellar systems which consist of two or more starlike objects) is
34.0\%, while the companion star frequency (or the number of pairs) 
is 44.7\%. Both values are very close to those derived by RG97.

\subsection {The mass-luminosity relation}

Computing the stellar mass function from these data demands that we adopt a
mass-luminosity relation. RG97 derived mass functions for their 8-parsec sample 
using Henry \& McCarthy's (1993 - HMc93) empirical (mass, M$_K$) relation, Kroupa et al's
(1993) semi-empirical relation and theoretical calibrations by Burrows et al (1993)
and Baraffe \& Chabrier (1996)\footnote{Note that RG97 calculations are made on a
star by star basis, not via a calculation of the nearby-star luminosity function}.
The results are in good agreement, save for the
calibration based on the (M$_{bol}$, mass) relation predicted by the last-mentioned set
of models. On that basis RG97 adopted the HMc93 empirical relation, derived from
astrometric orbit analysis or nearby stars, as their standard.

Baraffe et al (1998 - BCAH98) present colour-magnitude and mass-luminosity
relations for a revised set of solar-abundance models, and figure 1a compares their predictions
for ages of 0.1, 1 and 10 Gyrs 
against both the HMc93 calibration and available measurements of stellar masses
\footnote{The 'kink' in the empirical relation at 0.5M$_\odot$ reflects its
origins as two linear relations in the (M$_K$, log$M$) plane}.
Astrometric orbital determinations are supplemented by data for eclipsing binaries as
summarised by Andersen (1991), including YY Geminorum; analysis of the well-known
M4.5 binary CM Draconis (Metcalfe et al, 1996)\footnote {CM Draconis is
misplotted in figure 2 of Reid (1998).}; and data for the recently-discovered 
mid-M eclipsing system, GJ 2069A (Delfosse et al, 1999b). We have estimated M$_K$ for
both components in the last system based on the absolute visual magnitudes listed by
Delfosse et al and the (M$_V$, (V-K)) relation for nearby stars. Those estimates, to
which we assign generous uncertainties of $\pm0.2$ magnitudes, are consistent with
values based on the spectral type. Both stars are subluminous by $\sim1$ magnitude
as compared with the theoretical and empirical calibrations, as expected given the 
discrepancies noted by Delfosse et al at visual wavelengths, and adding to the 
already substantial dispersion in the diagram for $1.0 > {M \over M_\odot} > 0.3$. 
There is no evidence that the system is metal-poor based on the CaH and TiO
bandstrengths (Reid et al, 1995). Moreover, the
discrepancy is not restricted to astrometric binaries:
GJ 2069Ab has similar photometric properties to the CM Dra components, but almost twice the
mass. Clearly stars in this mass range, and GJ 2069Aab in particular, require more
attention, but for the present we must assume that we can derive a mass-luminosity
relation which is valid for the average disk dwarf. 

The location of the mean mass-luminosity relation is predicted to vary as a function 
of age and abundance, while the detailed shape is tied to the underlying
stellar physics, notably H$_2$ dissociation at $\sim0.7 M_\odot$ (Copeland et al, 1970)
and the onset of degeneracy below 0.15M$_\odot$ (Grossman, 1970). D'Antona \& Mazzitelli
(1985) originally emphasised the potential for systematic errors if $\Psi(M)$ is
derived by applying simplified  mass-luminosity relations to luminosity function
data. The importance of changes of slope in (L, $M$) depend on the number
of stars at the relevant luminosities. Rather than bin the data to give $\Phi$(M$_K$, we
plot the distribution of the 8-parsec sample in the M$_K$ vs r$^3$ plane (where r is the distance)
in figure 1b. This diagram permits a visual assessment of both the number density of
stars near points of inflection in the (L, $M$) relation, and of the completeness of
the sampleas a function of distance. The distribution of points suggests
a scarcity of very low-mass M dwarfs, M$_K > 9$, beyond $\sim 6$ parsecs.

Both the range of chemical abundances and the 
age distribution of the calibrating stars can influence the derivation of
$\Psi(M)$ by biasing the derived (mass, M$_K$) relation. In the former case, the
contribution to the dispersion in derived masses is likely to be small, since
Hipparcos colour-magnitude data indicate that $\sim90\%$ of disk dwarfs have
metallicities in the range $-0.3 < [M/H] < 0.1$ (Reid, 1999). Moreover, our
choice of M$_K$ in calibrating masses should minimise metallicity variations. 

Age variations are likely to affect only the lowest masses. The BCAH98 
isochrones plotted in figure 1 show no evidence for significant evolution in M$_K$
between ages of 1 and 10 Gyrs, but indicates that very low-mass ($<0.1 M_\odot$) 
dwarfs can be up to 1 magnitude brighter in K at age 0.1 Gyr than 1 Gyr. There
have been suggestions that the binary stars observed by HMc93 are biased toward
younger ages, most recently by Chabrier \& Baraffe (1998). Almost all  of those proposals
are based on the detection of chromospheric and/or coronal activity amongst the
lower-luminosity dwarfs in the HMc93 sample, and all are based on a misconception.
Chromospheric activity in M dwarfs is a function of {\sl both} age and {\sl mass}, and
persists for several Gyrs in mid- and late-type dwarfs. Hawley et al (1996)
have compiled statistics of activity as a function of type and show that over 50\%
of dwarfs with spectral types of M5.5 or later are dMe stars; Hawley \& Reid (1999) have
detected emission in an M5 dwarf in the 5 Gyr-old cluster M67; and, of
particular relevance in this case, fully 80\% of the 2MASS M7 to M9 dwarfs with spectroscopic
observations have substantial ($>5$ \AA\ equivalent width) emission (Gizis, in prep.). 
Thus, activity amongst the lowest luminosity dwarfs is {\sl expected} and cannot be
taken as a sign of youthfulness. 

Chabrier \& Baraffe comment that the three lowest mass HMc93 dwarfs 
are overluminous compared with their model predictions, again inferring this as
a sign of youthfulness. However, they explicitly plot the three dwarfs with the 
lowest measured masses. 
A symmetric distribution of observational errors demands that those objects will
also be overluminous. The full set of low-mass dwarfs is
well distributed about the mean relation plotted in figure 1, and similar
circumstances prevail for the extended sample of
low-mass binary components with orbits and mass determinations derived from
HST-FGS observations (Henry et al, 1999). The (mass, M$_V$) relation 
defined by those stars is identical to that deduced by Henry \& McCarthy (1993). Since {\it in toto}
these stars represent essentially all of the local binary systems accessible to astrometric 
observations, an unbiased sample, this internal consistency argues 
strongly that {\sl all} lie close to their main-sequence configuration.

\subsection {The mass function}

Defining our terminology, the stellar mass function is written as
\begin{displaymath}
\Psi(M) \qquad = \qquad {dN \over dM} \quad {\rm stars\ per \ unit \ mass}
\end{displaymath}
Following Salpeter's (1956) pioneering analysis, the mass function is
often expressed as a power-law. In that case it is useful to define
\begin{displaymath}
\xi(M) \qquad = \qquad { dN \over dlog(M)} \quad {\rm stars  \ per \ unit \ log (mass)}
\end{displaymath}
A power-law mass function $\Psi(M) \propto M^{-\alpha}$ corresponds to
$\xi(M) \propto M^{-\alpha + 1}$, also written as $\xi(M) \propto M^\Gamma$
(cf. Scalo, 1998). This formulation provides a convenient way of representing
$\Psi(M)$, which we adopt in this paper. This should not be misinterpreted
as a statement that $\Psi(M)$ {\sl is} a power-law. 

Figure 2 shows mass functions derived from data for the revised 8-parsec sample using
both the empirical calibration and the BCAH98 1 Gyr isochrone (adopting the 10 Gyr
isochrone does not change $\Psi(M)$ in any significant fashion). White dwarfs
are excluded. The two calibrations give results in good agreement,
with the empirical relation giving a somewhat
larger spread in mass below 0.1M$_\odot$. For present purposes we are concerned
with the general form of the mass distribution, rather than assessing whether any
indications of structure have statistical weight (the individual uncertainties
argue against that proposition) and fit a simple power-law over the 
1.0 to 0.1 M$_\odot$ mass range. The best-fit relations are
\begin{displaymath}
log \xi_*(logM) \qquad = \qquad -0.13\pm0.14 \quad log M \quad + \quad 1.02\pm0.08
\end{displaymath}
for the empirical (HMc93) calibration, and 
\begin{displaymath}
log \xi_*(logM) \qquad = \qquad 0.02\pm0.14 \quad log M \quad + \quad 1.10\pm0.08
\end{displaymath}
for the BCAH98 (mass, M$_K$) relation. In these equations
M is the mass in solar masses and the units are the number of 
stars per 0.1log(M) within the volume of the northern 8-parsec sample
(1608 pc$^{-3}$). These relations correspond to mass functions
$\Psi_*(M) \propto M^{-1.13\pm0.14}$ and $\Psi_*(M) \propto M^{-0.98\pm0.14}$ 
respectively, and a space density of 0.035 
stars pc$^{-3}$ (0.1 M$_\odot)^{-1}$ at M = 0.1 M$_\odot$.  

Nearly one third of the main-sequence dwarfs included in the 
star-by-star mass function are companions
of more massive stars. Excluding those stars (i.e. limiting the sample to single stars and 
primaries in multiple systems) gives the mass functions plotted as a dotted lines in figure 2.
These are determinations of the systemic mass function, $\Psi_{sys} (M) \propto M^{-0.92\pm0.13}$ from
HMc93 and $\Psi_{sys} (M) \propto M^{-0.89\pm0.13}$ is the results are based on the BCAH98 models. 
The corresponding space density of stellar {\sl systems} at 0.1 M$_\odot$ is 
0.02 systems pc$^{-3}$ (0.1 M$_\odot)^{-1}$, or $\sim60\%$ of the star-by-star space density. In
the simulations described in the following sections, the reference is set by
the mean density of stellar systems, rather than individual stars, in the 0.1 to 1 M$_\odot$ mass range.

\section {Modelling the local space density of brown dwarfs}

An accurate determination of $\Psi_* (M)$ at masses below 0.15 M$_\odot$
is rendered difficult by the low intrinsic luminosity of VLM dwarfs and
the consequent difficulties involved in finding such stars in even the
immediate vicinity of the Sun. VLM dwarfs, however,  at least achieve a stable, 
long-lived configuration in (log(L), T$_{eff}$), defining the main sequence in the HR diagram,
and leading to mass-luminosity and mass-effective temperature
relations which are effectively single-valued (cf. figure 1). The latter
situation does not prevail for brown dwarfs, which descend rapidly
through the HR diagram with $L \propto M^{2.6} t^{-1.3}$ and
$T_{eff} \propto M^{0.8} t^{-0.3}$ (Burrows \& Liebert, 1993). {\sl All}
substellar-mass objects evolve through at least part of the L-dwarf
sequence, albeit at very different rates. As a result, one cannot associate
a given spectral type (or luminosity) with a specific mass. Instead, we 
can only constrain $\Psi(M)$ at masses below 0.08 M$_\odot$ by comparing
the overall distribution of the observed characteristics against 
predictions based on theoretical simulations.

Evolutionary models of low-mass stars and brown dwarfs constitute the
backbone of our simulations. The two most recent sets of such models
are by Burrows et al (1997), who extend the calculations undertaken by
Burrows et al (1993), and by Baraffe et al (1998). The former set takes
account of opacities due to dust grains, which are predicted to form
at temperatures below $\sim2400$K, but use gray atmospheres to define the
outer boundary conditions; the latter employ the latest model atmospheres
computed by Allard \& Hauschildt, but include no grain opacities. Both 
predict similar isochrones, as figure 3 shows, with differences of only
100-200K at stellar masses. The Baraffe et al models predict 
similar luminosities but lower temperatures for young- and
intermediate-age low-mass stars, and higher luminosities and higher
temperatures at substellar masses (and for M$\le 0.08 M_\odot$ at age 10 Gyrs).
Given that we are concerned primarily with the latter
low-mass range, where grains make a substantial contribution to the overall opacity and
temperature structure, we have taken the Burrows et al models as the reference
for the current analysis. 

Having chosen a reference set of isochrones, our aim is to model the 2MASS and DENIS L-dwarfs -
magnitude-limited samples drawn from several-hundred square degree areas.
We have measurements of near-infrared apparent magnitudes
(IJK$_S$ (DENIS) and iJHK$_S$ (2MASS), where i is
determined from Keck spectra, see K2ML) 
and spectral types for each object, but
direct distance determination for only a handful. In contrast, the stellar models predict
intrinsic luminosity and effective temperature as a function of mass and
age. Given that our goal is to match both the number of L-dwarfs detected 
and the spectral-type distribution of those L-dwarfs, we require the following:
first, estimates of the stellar birthrate, B(t), and the initial mass function, 
$\Psi(M)$, to predict the local space density of low-temperature dwarfs; second, an
estimate of the K-band bolometric corrections to transform log(L) to M$_K$ and predict 
the number of sources detected by 2MASS; and, third, a relation between effective
temperature and spectral type, to estimate the distribution of detected sources. 
None of these relations is well established observationally. However,
the approximate L-dwarf effective temperature scale outlined in K2ML
provides the basis for a preliminary analysis of this issue.

\subsection {The spectral-type/effective temperature relation}

Effective temperature calibration for late-type dwarfs has been a subject of
some debate for well over a decade. Different analysis techniques have produced
temperature scales which disagree by 300K or more at spectral types later than
M7. Blackbody fitting to broadband photometry, originally applied by Greenstein, 
Neugebauer \& Becklin (1970), generally leads to the coolest temperature estimates,
while near-infrared and optical spectroscopic analyses, matched against 
stellar models, give progressively hotter scales. Thus, temperature estimates for
the well-known M8 dwarf VB10 have ranged from 
$\sim2330$K (Tinney et al, 1993a - blackbody fitting)
to $\sim2750$K (Jones et al, 1995 - near-infrared spectra) and $\sim2875$K (Kirkpatrick
et al, 1993 - optical and near-infrared spectra). 

Recent theoretical analyses have identified a different method of temperature
determination. Tsuji et al (1996a) originally suggested that dust grains, long 
known to be present in the low-density atmospheres of asymptotic giant
branch stars, should also form in cool main-sequence dwarfs. Dust formation
depletes the gas-phase population of a number of molecular species, leading
to significant changes in the emergent spectrum, notably the weakening and
subsequent disappearance of TiO and VO absorption. It is now clear that this
process is responsible for the hydride-dominated optical spectra of L dwarfs.

Tsuji et al (1996b) and Jones \& Tsuji (1997) have computed model atmospheres 
which include TiO depletion and find that these ``dusty" models give much improved
agreement between the predicted and observed depths of the near-infrared
water bands in late-type M-dwarfs. Their temperature estimates of $\sim2600$K
for Gl 406 (M6), $\sim2200$K for VB 10 and $\sim2000 - 2200$K for LHS 2924 (M9) are 
comparable with blackbody scales, even though the overall energy distributions of both
late-type M-dwarfs and L-dwarfs are far from Planck curves. Nonetheless, 
Tinney et al (1993a), anchoring their
blackbody curves using L-band (3.4$\mu m$) photometry,   
estimated T$_{eff} \sim 2080$K for 
LHS 2924 and T$_{eff} \sim 2600$K for Gl 406. 

Burrows \& Sharp (1998) have extended grain condensation calculations to include 
a wider range of molecular species, and K2ML demonstrate
that there is broad agreement between theoretical calculations of the expected
appearance and disappearance of different molecules with decreasing temperature 
and the observed
progression of bandstrengths in the L-dwarf sequence. Detailed analyses lead to
temperature estimates of $\sim1900$K for the L2 dwarf Kelu 1 (Ruiz et al, 1997),
and 1800-1900K for the L4 dwarf GD 165B (Tsuji et al, 1996b; Kirkpatrick 
et al, 1999b).
The decreasing strength of CrH and, perhaps, Li I, together with the absence of
the 2.2$\mu m$ band due to CH$_4$ suggest a temperature of between 1400 and 1500K
for the coolest-known L-dwarf 2MASS J1632291+190441 (type L8). 

Combining these results, we derive the approximate temperature scale plotted in figure 4.
The L-dwarf sequence is characterised as lying between effective temperatures of 
2000K (type L0/L1) and $\sim1400$K (i.e. somewhat cooler than spectral type L8). 
It is likely that there is some overlap between L0/1 dwarfs and the latest
M-dwarfs in the 2000-2100K temperature range. There are obviously appreciable uncertainties in
both the boundary values we adopt and in the overall scale, and those uncertainties are
taken into account in the analysis of the simulations described in the following
section.

\subsection {K-band bolometric corrections for L-dwarfs}

Transforming bolometric luminosities to M$_K$ magnitudes is a relatively straightforward
process. Figure 5 plots BC$_K$, where
\begin{displaymath}
BC_K \qquad = \qquad M_K \quad - \quad M_{bol}
\end{displaymath}
as a function of effective temperature (using the scale defined in the previous 
subsection) for a number of well-known dwarfs. As yet, GD165B and Gl 229B are the
only ultracool dwarfs with flux measurements at wavelengths longward of 2.2$\mu m$. However,
the peak in the emergent
spectral energy distribution lies close to the K-band over this 
range in temperature  and little variation is expected 
in the value of BC$_K$ from spectral type $\sim$M4 through L4 (GD 165B) to
$\sim$L8. At some temperature between the latter spectral type and the $\sim950$K
measured for Gl 229B, methane becomes a significant absorber at near-infrared wavelengths, 
removing $\sim60\%$ of the flux emitted in the K passband. Burrows and Sharp place
this transition at a temperature between 1500 and 1200K. To date, no brown dwarfs have been
discovered with properties intermediate between those of 2MASS J1632291+190441, which has
no significant methane absorption at 2.2$\mu m$, and the lone T-dwarf, 
Gl 229B, so it remains unclear both
exactly where methane absorption becomes important and whether the onset is gradual
or abrupt. 

The shape of the L/T transition in the (BC$_K$, T$_{eff}$) plane affects predictions
of both the colour distribution and the expected total numbers of detectable brown
dwarfs with T$_{eff} < 1400$K. Increasing BC$_K$ at a given temperature implies 
a fainter M$_K$ and hence a smaller spatial volume accessible to a magnitude-limited
survey. At the same time, an increased BC$_K$ leads to bluer (J-K) colours, since
the J passband is unaffected by CH$_4$ absorption (BC$_K \sim -3$ corresponds to
(J-K)$\sim 1.0$).
We have explored the different possibilities to a limited extent by considering
two ad hoc relations: an abrupt decrease in BC$_K$ of 1.3 magnitudes between
temperatures of 1400 and 1300K (relation A in figure 5); and a similarly sharp
transition between 1200 and 1100 K (relation B). These are chosen to span the
theoretically-predicted temperature range for the onset of substantial methane
absorption. Relation B predicts larger numbers of L dwarfs to a given apparent 
magnitude limit, but fewer 2MASS-detectable brown dwarfs with intermediate
or blue (J-K) colours.

\subsection {The stellar birthrate and $\Psi(M)$.}

The temperature distribution of brown dwarfs in the Solar Neighbourhood depends on 
both the mass and the age distribution, and is therefore tied directly
to the convolution of B(t) and $\Psi(M)$. However, the presently-available
observations do not allow us to separate the two functions. Figure 6 shows the
predicted evolution of effective temperature as a function of time for
objects with masses between 0.009M$_\odot$ and 0.1M$_\odot$. As discussed
by Burrows et al (1997), objects segregate into two well-defined
categories: those sufficiently massive to ignite central hydrogen-burning
and achieve a stable, main-sequence configuration with a near-constant
effective temperature, M$\ge 0.08M_\odot$; and brown dwarfs, $M < 0.07M_\odot$,
which show monotonic decrease in T$_{eff}$, after an initial deuterium phase
for M$\ge 0.015M_\odot$. VLM dwarfs with masses between $\sim0.075$ and 0.08M$_\odot$ are
transition objects, which maintain near-constant T$_{eff}$ for a substantial
fraction of a Hubble time.

Our best estimate of the L-dwarf domaine, $2000 > T_{eff} > 1400$K, is indicated
in figure 6. Given those temperature limits, 
M9 dwarfs such as LHS 2924 and BRI0021 are predicted to
have masses of $\sim0.08M_\odot$, with L0 dwarfs slightly less massive. Thus, based
on the Burrows et al (1993, 1997) evolutionary tracks, the only stars
expected to enter the L-dwarf r\'egime are those within $\sim$0.04M$_\odot$ of the
hydrogen-burning limit. All other L-dwarfs
are brown dwarfs.

Figure 6 illustrates an important characteristic of the L-dwarf population:
since L-dwarfs are defined by the effective temperature, our sample (indeed any sample) includes 
transition objects and brown dwarfs which not only span a wide range of age, but also
show little overlap in the age distribution at different masses. As an extreme
example, while a 0.075 M$_\odot$ brown dwarf enters the L-dwarf r\'egime at
an age of $\sim1.25$Gyrs and cools to 1400K at age $\sim10$Gyrs, a 0.009 M$_\odot$
object has a spectral type of L0 at age 3 Myrs and L9 at age 12 Myrs. This has
two important consequences: first, higher-mass brown dwarfs and VLM stars
make a proportionately larger contribution to the L-dwarf population - indeed, given
the absence of star formation in the immediate vicinity of the Solar Neighbourhood and
the limiting magnitude of 2MASS, there is very little chance of our identifying
any free-floating brown dwarfs below 0.015M$_\odot$;
second, short-term variations in the star-formation rate can mimic changes in the slope
of the initial mass function - and, conversely, we cannot infer B(t) from these
data without making assumptions about $\Psi(M)$.  

Other methods can be used to estimate B(t) for the Galactic disk, notably 
the distribution of chromospheric activity in the local stellar population.
These analyses are complicated to some extent by likely variations in the activity
of individual stars - the Maunder Minimum phenomenon in solar-type dwarfs. However,
Soderblom et al (1991) have demonstrated that the available data for G dwarfs and
M dwarfs are consistent with a uniform star formation rate over the past 9 Gyrs.
Henry et al (1996) have extended the observational sample to include some 800
G-type southern stars, and confirm the distribution of activity deduced from the
northern samples. Noh \& Scalo (1990) have used the white dwarf luminosity function to
probe the star formation history, and find some indications of
increased activity within the last 10$^9$ years, although the evidence is not conclusive. 
Given these circumstances, we assume a uniform star 
formation rate in the present simulations.

\subsection {The HR Diagram}

Having adopted an effective temperature scale and the relevant bolometric
corrections, we can compare the observed location of the four L dwarfs with
measured trigonometric parallax against the Burrows et al model predictions. 
Our temperature scale is tied to the Burrows \& Sharp phase-transition calculations, which
do not predict stellar luminosities, so this comparison represents an independent test 
of the consistency of the adopted temperature scale. Figure 7 plots representative 
evolutionary tracks from the Burrows et al models. Data for Gl 406
(M6), VB10 (M8) and LHS 2924 (M9)
are plotted as solid points, while the open circles mark the
location of (in decreasing luminosity) 2MASS J0345432+254023 (L0), 
2MASS J1439284+192915 (L1),
GD 165B (L4) and DENIS-P J0205.4-1159 (L7: Table 10 in K2ML). 
The two early-type L dwarfs are both more luminous than the slightly-hotter
star LHS 2924. These are the only two L-dwarfs in our sample with
I magnitudes brighter than 17th magnitude, and our Keck
HIRES echelle spectroscopy (Reid et al, in 
prep.) indicates that 2MASS J0345432+254023 is a double-lined spectroscopic
binary. 2MASS J1439284+192915 warrants further investigation. SB2 systems are
intrinsically more luminous than single objects, and one expects a
bias towards such systems in a magnitude-limited survey.

A striking characteristic of the model predictions is the small variation
in luminosity with changing mass at lower temperatures. This reflects the fact 
that radii are determined primarily by electron degeneracy and vary 
little as a function of mass. As a result, moderately accurate photometric
parallaxes can be calculated for brown dwarfs, regardless of the mass, once a
minimal grid of calibrators is established.

\section {Simulating the Solar Neighbourhood}

Both the DENIS and 2MASS consortia have constructed complete samples of
VLM dwarfs, drawn from initial subsets of their respective surveys.
Our aim is to create a computer model of the immediate environs of the Sun which
can be 'observed' in a manner which takes account of the biases inherent in
compiling those L-dwarf samples. We have used Monte Carlo techniques to 
simulate the (L, T$_{eff}$) distribution of Solar Neighbourhood brown dwarfs based
on different assumptions for the underlying mass function. We can then compare the
expected surface density of both L-dwarfs and methane dwarfs against our
observations. 

\subsection {Observational constraints}

The 2MASS L-dwarfs were identified through 
spectroscopic follow-up observations of
a complete magnitude-limited, colour-selected sample covering an area of 371 square
degrees - 0.9\% of the sky (full details are given in K2ML).
Seventeen of the 51 candidates prove to have spectral types
in the range L0 to L8. The principal selection criteria are K$_S \le 14.5$ (corresponding 
to a signal-to-noise limit of 10 for the lowest-sensitivity scans)
and (J-K$_S) \ge 1.3$. The photometric uncertainties 
in (J-K$_S$) vary from 0.05 to 0.15 magnitudes. Moreover, 
observations of other candidates (e.g. 2MASS J1439284+192915) show that the 
earliest-type L-dwarfs (L0, L1) can have (J-K$_S$) colours $\sim 0.15$ magnitudes 
bluer than the adopted colour limit. Thus, the sample is incomplete for 
spectral types earlier than $\sim$L2. 

An additional selection criterion is that there are no optical counterparts within
5 arcseconds of the current (2MASS) position on the USNO A catalogue (Monet et al, 1998).
For the northern celestial hemisphere, the latter is based on digitised scans of 
POSS I O (blue) and E (red) plates, including only those sources detected on {\sl both}
plates (limiting magnitude B$_O \sim 21$, R$_E \sim 20.5$).
Late-type M-dwarfs such as LHS 2065 (M9) have (R$_C$-K) colours
of at least 6.5 magnitudes, where R$_C$ is the Cousins R-band. However, the 
POSS I R$_E$ passband has half-power points at $\sim6200$ and $\sim6600$\AA\
(Minkowski \& Abell, 1963), omitting the red half of the R$_C$ passband. 
Bessell (1986) and Tinney 
et al (1993b) have shown that there is a significant colour term between 
the IIIaF photographic R-band (POSS II/UK Schmidt Telescope) and standard
Cousins photometry; the colour term is more pronounced for POSS I data, with
R$_E$ between 1 and 1.5 magnitudes fainter than R$_C$ for extremely red objects. 
Thus we expect (R$_E$-K) colours of $\sim$8 magnitudes for early-type L dwarfs,
and redder colours for cooler dwarfs, while
all VLM dwarfs have (B$_O$-R$_E$) colours of $\sim 2$ magnitudes.
It is therefore just possible that nearby (r$< 10$pc) early-type L-dwarfs may
have escaped our photometric selection process, although at such distances a
tangential velocity exceeding 6 km s$^{-1}$ would result in a positional
displacement of $>5$ arcseconds over 40 years and a mismatch between 
POSS I and 2MASS. In any event, the volume encompasses by such hypothetical
systems is less than 10\% of the total surveyed, and the overall statistics are
correspondingly little affected.

In comparison, the DENIS brown dwarf mini-survey (Delfosse et al, 1997) is
complete to K$_S$=13.5 (S/N=3) and covers an area of 240 square degrees. With
a smaller initial sample of candidates, spectroscopic observations extend to
sources with significantly bluer colours than the (J-K$_S$)=1.3 limit of 2MASS,
and the sample of 3 L-dwarfs identified can be regarded as complete. With a
brighter limiting magnitude (r$_{max}(DENIS) \sim {r_{max}(2MASS) \over \sqrt{2.5}}$)
and only 65\% of the areal coverage, the DENIS survey samples only $\sim15\%$ of
the volume covered by 2MASS. This is consistent with the relative number of 
L dwarfs found by the two surveys in their respective complete samples - three versus
seventeen. Based on these results, we can estimate the surface density of L-dwarfs with
spectral types of L2 or later and 
K$_S < 14.5$ as $0.032\pm0.010$ sq. deg.$^{-1}$. 

The spectroscopic data provide a further constraint on the mass distribution
of the L-dwarf population - the prevalence of lithium absorption. Magazzu et al
(1993) originally pointed out that brown dwarfs with masses below $\sim0.05-0.06 M_\odot$
are expected to have central temperatures which never rise above $\sim2.4 \times 10^6$K,
the critical threshold for lithium destruction. As a consequence, lower-mass brown dwarfs are 
expected to maintain atmospheric lithium abundances at primordial levels despite being
fully convective during their evolution through spectral-type M.
Higher-mass objects deplete lithium at varying rates. Indeed, the location of the 
lithium-depletion boundary has been used by Stauffer et al (1998) to estimate the age
of the Pleiades cluster. Of the various evolutionary calculations,  Burrows et al
(1997) predict that a 0.07M$_\odot$ 
brown dwarf reaches 99\% depletion by age$\sim200$Myrs (T$_{eff} \sim
2700$K, or spectral type M5), while a 0.06M$_\odot$ brown dwarf preserves two-thirds of the 
initial lithium abundance even at age 20 Gyrs (T$_{eff}\sim 600$K); Ushomirsky et al's (1998)
analytic calculations are in good agreement, although indicating a somewhat lower temperature 
($\sim2300$K) for the 1\% abundance point of a 0.07M$_\odot$ brown dwarf; Chabrier et al's
(1996) numerical models place the depletion/no depletion boundary closer to
0.055M$_\odot$ (see figure 2 in Tinney, 1998). 
Despite these quantitative differences, these
calculations are in qualitative agreement in predicting that brown dwarfs which
are capable of destroying lithium effectively complete that depletion before entering the
L-dwarf temperature r\'egime. Thus, 
the observed fraction of L-dwarfs with detectable lithium 6708\AA\ absorption provides
an indication of the relative proportion of high- and low-mass brown dwarfs in the sample.

Finally, we can set limits  on the surface density of cool T-dwarfs
similar to Gl 229B. With methane absorption removing a significant fraction of the flux
emitted in the H and K bands, these objects have near-infrared colours comparable with
A and F stars, but the steep spectrum at shorter wavelengths leads to extremely red optical/infrared 
colours. Thus, Gl 229B has (J-K$_S)\sim -0.1$ but (I-K$_S)\sim 6$ (Matthews et al, 1996).
We have now analysed $\sim1500$ square degrees of data from the 2MASS 
project for sources with JHK detections, J$ < 16.0$, (J-K$_S)<1.3$
and no optical counterpart on POSS I (Burgasser et al, in prep.),
searching both for T-dwarfs and for intermediate-coloured objects making the 
transition from spectral type L to T as the CH$_4$ 2.2 $\mu m$ bands develop. To date, we
have follow-up near-infrared observations of some 200 sources, including all of
those with (J-K$_S) < 0.9$, 96\% with (J-K$_S) < 1.0$ and 85\% with (J-K$_S) < 1.3$.

None of the observed sources survive as plausible T-dwarf candidates. 
Most of the (J-K$_S >0.8$) candidates are 
visible on the POSS II R- or I-band plates, indicating (R-K$_S$) colours
of no more than $\sim 5$, too blue for L/T transition objects. Spectroscopy
indicates that these are likely to be early- or mid-type M-dwarfs.
No infrared source is detected at the appropriate position in the remaining 
cases. Almost all of the latter lie within 20 degrees of the ecliptic and
are likely to be 2MASS observations of uncatalogued asteroids. 
Three sources with K$_S < 14.5$ (with (J-K$_S$)=0.92, 1.17 and 1.24) currently
lack observations and remain possible.
Irrespective of the true identity of these transient sources, we can state that 
the surface density of T-dwarfs with K$_S < 14.5$ is less than $2 \times 10^{-3}$ 
sq. deg$^{-1}$.

\subsection {The simulations}

We use Monte Carlo techniques to generate a catalogue of VLM ``stars" and
``brown dwarfs" with known distance, luminosity and effective temperature
and estimate the expected surface density of L dwarfs and methane dwarfs.
Since 2MASS samples a larger volume than DENIS, our simulations are scaled to
match those more sensitive survey limits. Lower-mass brown dwarfs have larger
radii and higher luminosities at a given temperature. 
Figure 7 shows that 0.015 M$_\odot$ brown dwarfs are predicted to have
luminosities of log${L \over L_\odot} \sim -3.5$ (M$_{bol} \sim 13.4$) at
temperatures of 2000K. Given $\langle BC_K \rangle = 3.4$, this
corresponds to M$_K \sim 10$, or, for K$_S = 14.5$, a
maximum distance of $\sim80$ parsecs. In our simulations, we therefore
generate 10$^7$ particles with a uniform density distribution (i.e. 
N(r)$\propto r^3$) over the distance range 0 to 80 parsecs.

We have chosen to represent the mass function as a power law, $\Psi(M) \propto
M^{-\alpha}$. In part, this reflects the fact that a power-law mass function
is a good match to $\Psi(M)$ over the range 1.0 to 0.1 M$_\odot$; in part,
our choice reflects computational convenience. 
Each source generated by the simulations has an associated mass, between
0.01 and 0.1 M$_\odot$ drawn from a power-law mass function
with $\alpha=0, 1, 2$, and an age, $\tau$, uniformly distributed
between 0 and 10 Gyrs (i.e. B(t) = constant). A log-normal mass function with a
maximum at 0.1M$_\odot$ predicts detection rates close to the  
$\alpha=0$ model. Given M and $\tau$, we
use the Burrows et al models to predict L and T$_{eff}$, which, using the (BC$_K$, T$_{eff}$)
relation and specified r, give M$_K$ and K. It is reasonable to
expect some cosmic scatter in the (M$_K$, T$_{eff}$) plane for brown
dwarfs of a given mass, and we have allowed for that by assuming a
dispersion of $\sigma_K = \pm0.15$ magnitudes in the bolometric correction relation.
This technique effectively incorporates malmquist bias in the simulations, and
we can determine the fraction of 
sources with K$_S< 14.5$ and $2000 > T_{eff} > 1400$K (2MASS-detected
L dwarfs), as well as the number of cooler sources (2MASS-detected
T dwarfs) with K$_S< 14.5$. 

These simulations can be scaled to the expected surface densities 
using volume densities derived from the Northern 8-parsec sample
discussed in section 2. That sample includes 121 stars in 91 systems with
masses in the decade between 1.0 and 0.1 M$_\odot$, corresponding to 
space densities of 0.075 stars pc$^{-3}$ and 0.057 systems pc$^{-3}$. 
On that basis, we expect space densities of 0.0075 `stars' pc$^{-3}$ and
0.0057 systems pc$^{-3}$ with masses between
0.1 and 0.01 M$_\odot$ for $\alpha = 0$; 0.075 pc$^{-3}$ and 0.057 pc$^{-3}$ for
$\alpha = 1$; and 0.75 pc$^{-3}$ and 0.57 pc$^{-3}$ for $\alpha=2$. 
We have chosen the volume densities predicted for systems as the reference 
point for the simulations.

One point which should be noted is that our simulations make 
no attempt to allow for binarism amongst the observed L-dwarf sample. Binary L-dwarfs 
are known
to exist - the Pleiades brown dwarf PPl 15 is binary (Basri \& Martin, 1998) and
will become an L-dwarf in $\sim1-2$ Gyrs, while, as mentioned above, there
is strong evidence that 2MASS J0345432+254023 is a spectroscopic binary - but
at present we have no statistics on their frequency. However, if the 
distribution is similar to late-type M-dwarfs, then the overall binary fraction is
$\sim30\%$, with approximately half of those systems having nearly-equal
mass (RG97). The latter systems have the most significant effect on the 2MASS
statistics, since they are detectable at distances larger by $\sim \sqrt{2}$
or a factor of three increase in volume. If $\sim15\%$ of L-dwarf systems
fall in this category, then the simulations will underestimate the 
observed numbers in a magnitude-limited sample by $\sim30\%$. 

\section {Results}

\subsection {The substellar mass function}

We have employed two methods in matching the predictions of our simulations
against the available data, restricting ourselves to the deeper 2MASS 
sample: a qualitative comparison of the spectral-type
distribution and the predicted temperature distribution; and quantitative 
comparison of the observed and predicted L-dwarf surface densities. 
Figure 8 presents the former comparison. All three models predict 
similar temperature distributions, with decreasing numbers with
decreasing temperature - while brown dwarfs spend proportionately
less time at these higher temperatures, the higher luminosities 
($L \propto R^2 T^4_{eff}$ and $R \sim constant$) leads to significantly
larger sampling volumes. The dotted and dashed histograms divide the
contributions at masses of 0.07, 0.055 and 0.03M$_\odot$. Note the
substantially increased fraction of low-mass brown dwarfs in
the $\alpha=2$ simulation.

Qualitatively, these predictions are in general agreement with the
observations, although the statistics are sparse. The scarcity of
L0 dwarfs amongst the 2MASS sources is due to the colour-selection criterion,
(J-K$_S)> 1.3$. Considering the results of the simulations, it is reasonable 
to infer that this criterion leads to incompleteness 
at the 10-20\% level in the 2MASS L-dwarf sample.

More quantitatively, Table 2 gives the predicted surface densities
in terms of the numbers expected for an all-sky survey  to
K$_S$ = 14.5 magnitude. The `observed' numbers are extrapolated from
our sample of 17 L-dwarfs. In section 3.1 we noted
that it is likely that spectral types L0 and L1 extend to temperatures above 
2000K, and our sample, although incomplete at those temperatures, probably
includes a few such dwarfs. We have taken this into account by assigning 
three of the L0/L1 dwarfs to the higher temperature range. On that basis,
we expect 2MASS to detect $\sim1550$ L dwarfs 
with $2000 > T_{eff} > 1500$K (types $\sim$L1 to L8)
over the whole sky. The three T$_{eff} > 2000$K
dwarfs imply all-sky detections of at least 300 such sources - clearly a 
lower limit, since we know from subsequent follow-up observations that L0
dwarfs can have colours of (J-K$_S) \sim 1.25$. Finally, we list upper
limits for the surface densities of L dwarfs with spectral types later than L8
and T dwarfs. 

The 2MASS detections predicted by the simulations are divided into four
temperature ranges designed to match these observational constraints,
with the numbers further subdivided to show the mass distribution.
The two uppermost r\'egimes, 
$1500 < T_{eff} \le 2000$K and $2000 < T_{eff} \le 2100$K, correspond to
$\sim$L1 to L8 dwarfs and $\le$L0 dwarfs. Lower
temperature dwarfs are grouped under two headings: ultracool L dwarfs, 
spectral types $\ge$L8; and methane T dwarfs. The temperature ranges spanned by
those two classifications depend on the temperature adopted for the
onset of significant methane absorption, i.e. the transition from BC$_K$=-3.3
to -2.1 magnitudes in figure 5. In the case of relation A, ultracool L dwarfs
have $1400 < T_{eff} < 1500$K, while T dwarfs have temperatures below 1400K; in
relation B, the temperature ranges are 1200 to 1500K and T$_{eff} < 1200$K 
respectively. As one would expect, the latter relation predicts fewer detectable
T dwarfs and a higher surface density of late-type L dwarfs. 

Simple inspection of Table 2 shows that the observed surface density lies between that
predicted for $\alpha = 1$ and $\alpha = 2$, with the steeper mass function 
predicting a larger contribution from low-mass brown dwarfs. The $\alpha=0$ model (and
the log-normal $\Psi(M)$) predict significantly fewer candidates than we observe.
Considering only L dwarfs
with T$_{eff} < 2000$K, approximately one-third 
are predicted to have masses below 0.055M$_\odot$ for $\alpha = 1$, while
the fraction is closer to two-thirds for $\alpha = 2$. These predictions 
can be compared with relative numbers of stars with and without detected Li 6708\AA\
absorption - five of seventeen L-dwarfs in the 2MASS sample. Finally, the
$\alpha=2$ model predicts that, depending on whether BC$_K$ relation A or B is
adopted, we should have detected between 16 and 4
methane-rich T-dwarfs within the 1500 square degrees so far surveyed, whereas the
cupboard is currently bare.

Taken together, these three comparisons {--} the total number of L-dwarfs, 
of T-dwarfs and of lithium detections {--} suggest that if the mass function
is to be parameterised as a power-law, the index lies closer to $\alpha=1$
than $\alpha=2$, with $\alpha$ probably closer to the lower value. 
The data suggest a somewhat steeper slope
than the $\Psi(M) \propto M^{-0.6\pm0.15}$ derived by Bouvier et al (1998) in the
most recent analysis of low-luminosity Pleiades members, the only
other extensive sample of substellar-mass objects.
Our result is not strongly inconsistent with the 
index derived for the main-sequence stars (M$> 0.1 M_\odot$) in the 8-parsec sample 
and may indicate a degree of continuity across the hydrogen-burning limit.  
If we take $\Psi(M) \propto M^{-1.3}$ as a representative solution, then the 
predicted space density of brown dwarfs in the range 0.075 to 0.01 M$_\odot$ is 
$\sim0.10$ systems pc$^{-3}$, almost twice the density of 
0.057 systems pc$^{-3}$ derived for Solar Neighbourhood
0.1 to 1.0 M$_\odot$ main-sequence systems. Under these circumstances the average inter-system 
separation (stars and brown dwarfs) is $\sim1.9$ parsecs.

The expected number of T dwarf detections is strongly dependent on the 
CO/CH$_4$ transition temperature. With the onset of strong methane absorption, the
integrated flux emitted within the 2.2$\mu m$ window drops by almost a factor of 4, with
a corresponding decrease of a factor of eight in the volume accessible to a magitude-limited
survey. In contrast, the 1.2$\mu m$ J passband is little affected {--} Gl 229B has an almost
identical bolometric correction, BC$_J \approx 2$, as the M6 dwarf Gl 406 (Wolf 359). Since
the 2MASS survey has a limiting magnitude (10$\sigma$) of 16, this passband represents the 
optimum method of searching the database for candidate T dwarfs. Table 3 lists the 
predicted surface densities for $\alpha=1$ and $\alpha=2$ mass functions. Even the least
optimistic circumstance, $\alpha=1$ and the CH$_4$ transition at 1200K, predicts one 
detectable T dwarf per 500 square degrees, and the contrast between $\alpha=1$ and
$\alpha=2$ is substantial. Follow-up observations of a J-magnitude limited sample of 
2MASS sources are currently being undertaken (Burgasser et al, in prep.). 

\subsection {A comparison with previous surveys}

Our analysis indicates a substantial local space density of substellar-mass
objects. Given that conclusion, is it surprising that field brown dwarfs
were not discovered in previous surveys for low-mass stars? The simple answer is
no - the low luminosities and low temperatures lead to the overwhelming majority
of objects having extremely faint magnitudes at optical wavelengths. All of
the known L dwarfs are several magnitudes fainter than the m$_{pg}\sim21$ limit 
of the POSS I blue plates.  
A handful of the brightest 2MASS sources are barely visible in visual scans of
the POSS I E plates (m$_r$(lim)$\sim20.5$ to 21), although below the threshold 
for reliable detection in
automated scans\footnote{For some sources visual identification is aided by the 
knowledge that there {\sl is} a source near the given position.}.
In the few cases where comparison of POSS I and 2MASS astrometry is possible it is
clear that the  derived proper motion is more than 0.2 arcseconds per year, 
exceeding the criterion for Luyten's Two-Tenths proper motion survey (the NLTT). 
However, Luyten's 
second-epoch E-band plates have exposures times of only $\sim15$ minutes, significantly
shorter than the 40 minutes of POSS I (see the discussion in Reid, 1997), which leads 
to an effective limiting magnitude of m$_r \sim 19.5$ and eliminates all of the 2MASS
and DENIS sources and even the bright L-dwarf, Kelu 1. 
 
Almost all previous photometric wide-angle ($> 10$ square degrees) surveys
are based on
photographic material and required detections in both the R and I passbands\footnote
{ This
criterion is eminently sensible given the large numbers of spurious images 'detected'
on scans of photographic plates, especially IVN-emulsion I-band plates.}, with typical
limiting magnitudes of R$\sim20$ to 21 and I$\sim17$ to 18. The main exception is
Kirkpatrick et al's (1994) 27.3 sq. deg. CCD survey, which is complete to the
relatively bright magnitude of R$\sim 19$, although a few relatively
bright (I$< \sim18$) I-only sources were also observed spectroscopically.  
Our simulations, based
on (uncertain) I-band bolometric corrections, predict all-sky detection of
only a few tens of sources with I$<$17, and those numbers are consistent with  
the available observations. Considering only the deeper
2MASS survey, only two L dwarfs have I$< 17$ mag., implying a surface density
of no more than 1 source per 180 square degrees; five further sources have 
17$< I < 18$ (1 source per 74 sq. deg.), but all have {\sl photographic} R magnitudes
fainter than 21. Tinney's (1993) POSS II/UKST photographic survey is therefore the only
previous analysis which covers sufficient area (270 sq. deg.) to sufficient depth in
the I-band to have detected even a few field L dwarfs, but the requirement for (R-I)
colours in that survey
likely eliminated any such sources from further examination. 

This comparison underlines a key component of the initial justification for 
2MASS and DENIS:  previous surveys based on wide-field optical photometry {\sl are}
poorly suited to detecting brown dwarfs and can offer only very weak constraints 
on $\Psi(M)$ at substellar masses. Indeed, even the 2$\mu m$ surveys currently underway
are capable of detecting only a relatively small fraction of the substellar-mass
populations hypothesised in this paper. Figure 9 plots the predicted luminosity
functions, $\Phi$(M$_K$) and $\Phi$(M$_{bol}$), and temperature distribution for 
the constant birthrate/($\alpha$=1,2) models discussed above. The numbers are
scaled to match an 8-parsec radius spherical volume centred on the Sun. Even with
the flatter mass function, $\alpha=1$, almost 70\% of the population are
expected to have absolute magnitudes M$_K > 16$ and therefore are not detected by 2MASS.

Finally, Figure 6 shows that low-mass brown dwarfs spend a modest, but appreciable,
fraction of their lifetimes at temperatures above 2000K. As discussed in
section 4.1, those should be the only objects with a significant atmospheric
abundance of lithium at temperatures below $\approx 2400$K. Thus the
observed frequency of lithium detections amongst late-type M dwarfs can provide
another constraint on $\Psi(M)$. Given a lithium-destruction threshold at
0.06 M$_\odot$ and a volume-limited sample of dwarfs with temperatures in
the range 2400 to 2000K ($\approx$M7.5 to L0), our simulations predict that
one-sixth should have essentially undepleted lithium for $\Psi(M) \propto M^{-1}$; the
fraction rises to one-third if $\Psi(M) \propto M^{-2}$. 

Observational samples of late-type M-dwarfs remain relatively sparse: Kirkpatrick et al's
list of VLM dwarfs in the Solar Neighbourhood includes 16 stars with spectral types
in the range M7.5 to L0, while the APM sample (Kirkpatrick et al, 1997a) includes
an additional two stars; follow-up spectroscopy of the 2MASS prototype camera
yielded six dwarfs (Kirkpatrick et al, 1997b); and Tinney et al (1998) classify five
of the DENIS brown dwarf mini-survey candidates as type M8 and M9. Amongst that sample of
29 dwarfs, only one, LP 944-20 is confirmed as a brown dwarf through the
detection of the lithium 6708\AA\ line at an equivalent width of 0.5\AA\ (Tinney, 1998) -
a significantly smaller fraction than even the predictions of the $\Psi(M) \propto M^{-1}$ 
model. However, lithium
was {\sl not} detected by Kirkpatrick et al (1997a), whose observations of LP944-20 were made
at a lower resolution (8\AA\ ) for the purpose of spectral classification. 
Indeed, all of the classification observations were made at spectral
resolutions of between 7 and 18 \AA\, and high signal-to-noise spectra at a 
resolution capable of detecting lithium 6708\AA\ are available for no 
more than ten of the twenty-nine dwarfs. Further observations of these objects, 
together with data for more extensive samples of VLM dwarfs culled from the 2MASS and
DENIS surveys, will provide a substantive test of the models and conclusions 
of the current paper. 

\subsection {Brown dwarfs in binary systems}

Both GD165B and Gl 229B, the latter the archetype (indeed the only) T-dwarf, were
identified in the course of surveys for brown dwarfs as companions to main-sequence
or post-MS stars. Some models for binary formation (e.g. capture processes) predict
that both components are selected at random from the same mass function. If both
components are selected from a mass function $\Psi(M) \propto M^{-1.3}$, then
in a sample of binaries with main-sequence primaries, approximately 
two out of three secondaries should be a brown dwarf. In contrast to 
that expectation, very few such systems have 
been discovered either by direct imaging (Rebolo et al, 1998; Oppenheimer et al, 1999) or in
the course of 
radial velocity surveys (Marcy \& Butler, 1998). The latter observations are most
sensitive to companions at small separations and short periods, and are
obviously not capable of detecting Gl 229B analogues. Might a substantial
number of the latter type of system remain undetected through the brown
dwarfs having faded to luminosities below detection limits of current 
surveys for wide binaries? We can address
that question using our simulations. 

Assuming a constant stellar birthrate and $\Psi(M) \propto M^{-1.3}$ for 
wide ($>30$ a.u.) main-sequence/brown dwarf  binary pairs, 
our calculations indicate that at distances of up to 8 parsecs,
$\sim15\%$ of companions with 0.075$\ge {M \over M_\odot} \ge 0.01$ 
have apparent magnitudes brighter than K$\sim 14.5$. Such objects should be
detectable in the deep imaging surveys undertaken by Oppenheimer et al (1999)
and Simons et al (1996), but Gl 229B remains the only detection. It is only
possible to both increase the frequency of brown dwarfs companions {\sl and}
match the observational constraints if $\Psi_C (M)$ is significantly steeper
than the mass function we deduce for the field. We conclude that the $\sim 1\%$
observed frequency of binaries with brown dwarf companions, as compared
with the $\sim25\%$ main-sequence systems with M-dwarf companions, indicates
that $\Psi_C (M)$ is not identical with the mass functions of isolated
stars and brown dwarfs.

\subsection {Mid-infrared surveys}

Theoretical models of low-temperature ($<$1400K) brown dwarfs predict that
a relatively high fraction of the bolometric flux is emitted at
wavelengths between 5 and 20 $\mu m$. The 
Wide-field Infrared Explorer (WIRE) satellite is scheduled to
undertake the Moderate-Depth Survey (MDS), 
covering between 800 and 1000 square degrees to
a (S/N=5) flux limit of 1.5 and 0.53 mJy at 12 and 25 $\mu m$ respectively
\footnote{ At the time of revising this paper, WIRE appears to be in critical condition.
We leave the calculations intact as an illustration of the
potential T dwarf harvest at mid-infrared wavelengths}. 
This represents an increase in sensitivity over previous observations at those wavelengths
(IRAS, ISO) comparable to 
the increment between the TMSS and 2MASS. WIRE offers the possibility of testing
our predictions of the space densities of substellar-mass objects, particularly 
at the lower end of the 0.07 to 0.01 M$_\odot$ mass range. 

The main limitation in extending  our calculations to longer wavelengths 
stems from uncertainties in the 12$\mu m$ bolometric corrections, BC$_{12}$.
Neither observation nor theory offers accurate definition of the variation
with T$_{eff}$, especially for
temperatures between $\sim2200$ and $\sim1300$K (i.e. spanning the full L-dwarf
sequence) where silicate dust formation is likely to influence the 
mid-infrared flux distribution.
 The next generation of VLM/brown dwarf models
(Burrows et al, in prep.) will be better placed to cope with these predictions.
For the present, we have defined the BC$_{12}$(T$_{eff}$) relation using two
datapoints: Gl 406 (M6, T$_{eff}\sim 2700$K) and Gl 229B. The former is the
lowest luminosity dwarf with a direct flux measurement at 12$\mu m$.
(by D. Aitken and P. Roche, reported in Berriman \& Reid, 1987) 
The flux density of 200 mJy corresponds to an absolute magnitude  M$_{[12]}$=8.4
\footnote{ In the [12] IRAS system, magnitude 0 is defined by a 28 Jansky source.
The WIRE MDS limit is [12]$\sim 11.8$. }.
The corresponding (K-[12]) colour is 0.7 mag. and BC$_{12} \sim 3.7$ magnitudes.
Gl 229B, on the other hand, has not been detected at 12$\mu m$, although Noll et al 
(1997) measure F$_\nu = 3.6\pm0.2$ mJy. at 5 $\mu m$. The Burrows et al (1997) predicted
flux distribution for a 1000K brown dwarf is in reasonable agreement with the latter
observation, and that model predicts F$_\nu \sim 2.3$mJy. for Gl 229B at 12$\mu m$
(K-[12]$\sim 4.1$ and BC$_{12} \sim 6.2$ magnitudes. Lacking any other
constraints at present, we assume a linear relation in BC$_{12}$ anchored at these
two points.

Bearing in mind the substantial caveat implied by these significant
uncertainties, we have calculated the expected surface densities of WIRE-detectable
L- and T-dwarfs. The expectations for a 1000-square degree WIRE MDS are shown in Table 4
for the $\Psi(M) \propto M^{-1}$ and $M^{-2}$ models {--} $\Psi(M) \propto M^0$
predicts a negligible rate of source detection. As in Table 2, we show the approximate
mass range of the predicted detections. The increasing bolometric corrections at
lower temperatures leads to a higher proportion of T-dwarf detections than in 2MASS,
and a greater contribution from lower-mass brown dwarfs. However, the extremely
low intrinsic luminosities of those objects means that the total expected
number of detections is small {--} $\approx 30$ for $\alpha=2$, and barely a handful
for $\alpha = 1$. All ought to be detected by 2MASS (at J and H, if not K), so the
red (K-[12]) colours should permit identification and provide additional
information on $\Psi(M)$ at masses below 0.05 M$_\odot$.

\subsection {The mass density}

The local mass density is little affected by the addition of these
brown dwarf systems to the Solar Neighbourhood census. An $\alpha=2$ model
places almost the same total mass in $M > 0.01 M_\odot$
brown dwarfs as in hydrogen-burning stars, but also predicts an L-dwarf
detection rate four times higher than the observed value. 
If $\Psi(M) \propto M^{-1.3}$, brown
dwarfs in the range 0.075 to 0.01 M$_\odot$ contribute only $\sim$0.005 
M$_\odot$ pc$^{-3}$, or
$\sim15\%$ of the mass density known to be due to main-sequence stars. Even if this 
mass function were extrapolated a further decade to 0.001M$_\odot$, the
resultant total mass density is only twice that due to
the nine white dwarfs known in the northern 8-parsec sample. Brown dwarfs
are the lumpen proletariat of the Solar Neighbourhood {--} ubiquitous, 
but with no visible influence.

\section {Summary}

We have presented a preliminary analysis of the form of the mass function
below the hydrogen-burning limit using initial results from
the 2MASS and DENIS brown dwarf projects. Both surveys have succeeded
in detecting objects with effective temperatures significantly cooler than
the latest-type M-dwarfs, with the deeper 2MASS data contributing 17 L-dwarfs
within a solid angle of 371 square degrees. Neither survey, however, has yet
identified field counterparts of Gl 229B, brown dwarfs with effective
temperatures cooler than 1200K and methane-dominated spectra. 

Combining these observational constraints with the frequency of 
lithium detection amongst the 2MASS L-dwarfs, we have used the suite of low-mass
star/brown dwarf models computed by Burrows et al (1993, 1997) to simulate
the Solar Neighbourhood, taking the stellar mass function derived from
stars within 8-parsecs of the Sun as the reference zeropoint. The
uncertainties inherent in such calculations are considerable, particularly
given the problem of transforming from the theoretical to the observational
plane, and the degeneracy between $\Psi(M)$ and the stellar birthrate
in defining the present-day luminosity and temperature distributions.
Nonetheless, the 
resulting predictions are consistent with the observations for a power-law
mass function with $1 < \alpha < 2$. The results are insensitive to the 
presence of binaries, povided that the L dwarf binary distribution
is similar to that observed for M dwarfs. 

If $\alpha = 2$, brown dwarfs with masses exceeding
0.01$M_\odot$ have the same mass density as hydrogen-burning stars, but 
the predicted surface density of 2MASS-detectable L dwarfs exceeds the observations
by a factor of 3.  For $\alpha = 1.3$, which is
more likely to be an appropriate value, brown 
dwarfs with masses exceeding 0.01M$_\odot$ outnumber stars by almost
a factor of two, but contribute only one-sixth the mass density. 
Based on these results, brown dwarfs are very unlikely to provide a significant contribution to any 
dark matter component within the Galactic Disk.

\acknowledgements
{JDK, INR and JL acknowledge funding through a NASA/JPL grant to 2MASS Core Project
science. AB acknowledges funding support from NASA grants NAG5-7073 and NAG5-7499. 
Thanks to Ben Zuckerman for prompting the calculation of lithium statistics for
late M-dwarfs. This publication makes use of data rom the 2-MASS All-Sky Survey, which is
a joint project of the University of Massachusetts and the Infrared Processing and
Analysis Center, funded by the National Aeronautics and Space Administration and the
National Science Foundation.}

{}

\clearpage

\begin{deluxetable}{crccrccl}
\tablewidth{0pt}
\tablenum{1}
\tablecaption{Revisions to the Reid/Gizis (1997) northern 8-parsec sample}
\tablehead{
\colhead{Name} & \colhead{M$_V$} & \colhead{(B-V)} &
\colhead{(V-I)} &\colhead{HIP} &\colhead{$\pi$} & \colhead{Comments} }
\startdata
  Rejected stars & \\
 & \\
Gl 178 & 3.67 & 0.46 & 0.52 & 22449 & 0.1246$\pm$0.0010 &Hipparcos $\pi$\\
Gl 190 & 10.45 & 1.53 & 2.64 & 23932 & 0.1073$\pm$0.0020&Hipparcos $\pi$\\
Gl 623A& 10.98 & 1.50 & 2.30 & 80346 & 0.1243$\pm$0.0016&Hipparcos $\pi$ \\
Gl 623B& 16.04 & \\
Gl 686 & 10.07 & 1.53 & 2.11 & 86287 & 0.1230$\pm$0.0016&Hipparcos $\pi$\\
Gl 713A & 4.04 & 0.49 & & 89937 & 0.1241$\pm$0.0005 &Hipparcos $\pi$\\
Gl 713B & 4.65 &\\
 & \\
LP 476-207AabB & 13 & & & & 0.11$\pm$0.03 & new component$^{1,2}$, $\pi_{spec}$ \\
G 89-32AB     & 13 & & & & 0.10$\pm$0.03 & new component$^1$, $\pi_{spec}$ \\ 
& \\
Additional stars & \\
& \\
Gl 382 & 9.80 & 1.50 & 2.17 & 49986 & 0.1280$\pm$0.0015 & Hipparcos $\pi$\\
Gl 701 & 9.91 & 1.50 & 2.06 & 88574 & 0.1283$\pm$0.0014 & Hipparcos $\pi$\\
LP 816-60 & 12.7 & & & 103039 & 0.1822$\pm$0.0037& Hipparcos $\pi$ \\
Gl 793 & 11.05 & 1.56 & 2.43 & 101180 & 0.1256$\pm$0.0011 &Hipparcos $\pi$ \\
& \\
Gl 829B & 11.9 & & & 106106 & 0.1483$\pm$0.0019 & new component$^2$\\
Gl 831C & 15.4 & & & & 0.126$\pm$0.023 & new component$^1$ & \\
Gl 896Aa & $\sim13$ &  & & 116132 & 0.1601$\pm$0.0028 & new component$^{2,3}$\\
LTT 1445C & $\sim14.5$ && &        & 0.127$\pm$0.025 & new component$^3$, $\pi_{spec}$ \\
\enddata
\tablecomments{
The first group of stars listed are rejected because the Hipparcos parallax 
is less than 0.125 arcseconds. \\
LP 476-207 and G89-32 are eliminated due to the smaller
inferred spectroscopic parallax. \\
Additions to the 8-parsec sample stem either from improved 
parallax data from the Hipparcos satellite or from new discoveries
in surveys for multiple stars - references:\\
1 - Henry et al, 1997 \\
2 - Delfosse et al, 1999 \\
3 - Oppenheimer et al, 1999 {--} LTT 1445 = LP 771-95/96 (RG97)}
\end{deluxetable}

\begin{deluxetable}{crcccccl}
\tablewidth{0pt}
\tablenum{2}
\tablecaption{Predicted L/T-dwarf surface densities - K$_S < 14.5$}
\tablehead{
\colhead{Temperature} & \colhead{total} & \colhead{M$>$0.07M$_\odot$} &
\colhead{0.07 $> {M \over M_\odot} > 0.055$} &
\colhead{0.055 $> {M \over M_\odot} > 0.030$} &
\colhead{0.03 M$_\odot$ $>$ M}} 
\startdata
 & && Observed &\\
T dwarf & $<82$ & \\
$\le$L8 &$<220$  &\\
1500 - 2000 & $\sim 1550$ &$\sim$L1 to L8&\\
2000 - 2100 & $>300$ &$\le$L0\\
 & \\
 & &&$\alpha = 0$ &\\
BC$_K^A$: T dwarf & 5.5 & 1.3 & 2 & 1.7 & 0.5 &\\
 $\le$L8 dwarf & 9 & 3.5 & 2.7 & 2.2 & 0.6&\\
BC$_K^B$: T dwarf &2.4 & 0.3 & 1 & 0.8 & 0.3 &\\
 $\le$L8 dwarf & 18 & 6.5 & 6.1 & 4.3 & 1.6&\\
1500 - 2000 & 90 & 58 & 15 & 12 & 5 &\\
2000 - 2100 & 32 & 24 & 4 & 3.5 & 1.5 & \\
& \\
 & &&$\alpha = 1$ &\\
BC$_K^A$: T dwarf & 64 & 10 & 18 & 22 & 14 &\\
 $\le$L8 dwarf & 44 & 7 & 12 & 16 & 9 &\\
BC$_K^B$: T dwarf & 17 & 2 & 5 & 6 & 5 &\\
 $\le$L8 dwarf & 154 & 34 & 41 & 48 & 31 &\\
1500 - 2000 & 607 & 300 & 91 & 108 & 108 &\\
2000 - 2100 & 206 & 123 & 24 & 31 & 27 &\\
& \\
 & &&$\alpha = 2$ &\\
BC$_K^A$: T dwarf & 435 & 34 & 59 & 119 & 223 &\\
 $\le$L8 dwarf & 434 & 50 & 54 & 107 & 223 &\\
BC$_K^B$: T dwarf & 119 & 1 & 23 & 38 & 57 &\\
 $\le$L8 dwarf & 1077 & 126 & 161 & 299 & 491 &\\
1500 - 2000 & 3653 & 929 & 334 & 648 & 1742 &\\
2000 - 2100 & 981 & 361 & 86 & 164 & 370 &\\
\enddata
\tablecomments{
Predicted detections of L- and T-dwarfs by the full (i.e. all-sky) 2MASS survey.
The observed numbers are derived by extrapolating from the 17 L-dwarfs detected within
371 sq. deg. (0.9\% of the sky) and from the three remaining (as-yet unconfirmed
T-dwarf candidates within a 1500 sq. deg. region (3.6\% of the sky). The upper limit listed for
ultracool L dwarfs (Sp $>$L8) corresponds to a surface density of two
objects within the area surveyed to date. The predicted detection
rates from each simulation are listed for comparison, with the expected numbers
of ultracool L dwarfs and T dwarfs computed using both hypothetical BC$_K$ relations 
(A and B) plotted in figure 5. The relevant temperature ranges are discussed in the text.}
\end{deluxetable}

\pagestyle{empty}
\begin{deluxetable}{crcccccl}
\tablewidth{0pt}
\tablenum{3}
\tablecaption{Predicted L/T-dwarf surface densities - J$< 16$}
\tablehead{
\colhead{Temperature} & \colhead{total} & \colhead{M$>$0.07M$_\odot$} &
\colhead{0.07 $> {M \over M_\odot} > 0.055$} &
\colhead{0.055 $> {M \over M_\odot} > 0.030$} &
\colhead{0.03 M$_\odot$ $>$ M}} 
\startdata
 & \\
 & &&$\alpha = 1$ &\\
$< 1200 $ & 81 & 3 & 23 & 33 & 22&\\
1200 - 1400 & 100 & 19 & 26 & 34 &21 &\\
1400 - 1500 & 61 & 16 & 14 & 18 & 13 &\\
1500 - 2000 & 629 & 314 & 92 & 114 & 109&\\
2000 - 2100 & 283 & 168 & 32 & 50 & 33 &\\
& \\
 & &&$\alpha = 2$ &\\
$< 1200$ & 577 & 13 & 89 & 198 & 278&\\
1200 - 1400 & 756 & 64 & 104 & 210 & 378 &\\
1400 - 1500 & 388 & 37 & 61 & 90 & 199&\\
1500 - 2000 & 3762 & 946 & 320 & 679 & 1817&\\
2000 - 2100 & 1362 & 475 & 132 & 272 7 482 &\\
\enddata
\tablecomments{
Predicted detections of L- and T-dwarfs for an all-sky survey to 16th magnitude in
the 1.2$\mu m$ J passband. }
\end{deluxetable}

\begin{deluxetable}{crcccccl}
\tablewidth{0pt}
\tablenum{4}
\tablecaption{Predicted L/T-dwarf surface densities - WIRE MDS}
\tablehead{
\colhead{Temperature} & \colhead{total} & \colhead{M$>$0.07M$_\odot$} &
\colhead{0.07 $> {M \over M_\odot} > 0.055$} &
\colhead{0.055 $> {M \over M_\odot} > 0.030$} &
\colhead{0.03 M$_\odot$ $>$ M}} 
\startdata
 & \\
 & &&$\alpha = 1$ &\\
$< 1400$ & 2.7 & 0.2 & 0.7 & 1.1 & 0.7 & \\
1400 - 1500 &0.3 & 0.1 &0.1  & 0.06 & 0.04 & \\
1500 - 2000 &0.6  & 0.3 & 0.1 & 0.1 & 0.1 &\\
2000 - 2100 & 0.4 & 0.2 & 0.05 & 0.1 & 0.05 &\\
& \\
 & &&$\alpha = 2$ &\\
$< 1400$ & 22 & 0.6 & 2.2 & 6.2 & 13 &\\
1400 - 1500 & 2.0 & 0.2 & 0.3& 0.6 & 0.9 &\\
1500 - 2000 & 3.7& 0.9 & 0.5 & 0.6& 1.7 &\\
2000 - 2100 & 2.9 & 1.0 & 0.2& 0.4 & 1.3 &\\
\enddata
\tablecomments{
Predicted detections of L- and T-dwarfs in the 12$\mu m$ Moderate-Depth  WIRE survey.
The predictions are for an area of 1000 square degrees 
surveyed to [12]=11.8 magnitudes. As noted in the text, 12$\mu m$ bolometric corrections
are {\it extremely} uncertain for sources in the L-dwarf/late-type M-dwarf temperature 
range.}
\end{deluxetable}

\clearpage
\centerline{FIGURE CAPTIONS}
\vskip2em

\figcaption{Upper:The (M$_K$, mass) relation: solid points are eclipsing binaries; open
triangles are from Henry \& McCarthy (1993). The solid line is the empirical
relation from the latter paper, while the dotted lines are theoretical relations
from Baraffe et al (1998) for ages of 0.1, 1 and 10 Gyrs.
Lower: the distribution of the 8-parsec stars in the (r$^3$, M$_K$)plane. Solid
points mark the brightest stars in individual systems.}

\figcaption{The stellar mass function defined by the northern 8-parsec sample.
The upper diagram plots the results if we adopt the empirical HMc93 calibration; the lower
for masses calibrated by the 1 Gyr BCAH98 isochrone. In both cases the
solid histogram/line delineates the star-by-star function; the dotted line
marks the systemic function.}

\figcaption{A comparison between the Burrows et al (1993, 1997) (solid squares) and Baraffe et
al (1998) models (open triangles) in the (log(L), T$_{eff}$) plane. The Burrows et al data 
are plotted for masses of 0.02, 0.03, 0.04, 0.05, 0.06, 0.07, 0.075, 0.08, 0.09, 0.1, 0.11 and 
0.15 M$_\odot$; the Baraffe et al datapoints are for masses of 0.035, 0.045, 0.05, 0.06, 0.070,
0.072, 0.075, 0.08, 0.09, 0.10, 0.13, 0.15, 0.175, 0.20 M$_\odot$ (0.1 Gyrs);
0.06 M$_\odot$ et seq. for 1 Gyr; and 0.075 M$_\odot$ et seq. for 10 Gyrs.}

\figcaption{ The effective temperature/spectral type relation adopted in this paper.}

\figcaption { K-band bolometric corrections as a function of temperature. The solid points
 are measured bolometric corrections by Tinney et al (1993a - Gl 406, VB10, LHS 2924 and
GD 165B) and Matthews et al (1996 - Gl 229B). The open symbols outline the 
 two relations adopted in our simulations to estimate the variation in BC$_K$ with the
 onset of CH$_4$ absorption: relation A is identified by open circles, relation
B by open triangles.}

\figcaption{ The effective temperature/age relation defined by the Burrows et al (1997, 1993)
models. The evolutionary tracks are plotted for masses of between 0.009 and 0.10 M$_\odot$, with 
the individual curves identified on the diagram. The horizontal lines outline the 
temperature range which correspond to the L-dwarf spectral type.}

\figcaption{ The HR diagram from VLM/brown dwarfs: we plot Burrows et al (1993, 1997) models 
for masses of 0.015, 0.03, 0.05, 0.06, 0.07, 0.08, 0.09 and 0.1 M$_\odot$ - the 
lowest-mass track lies at the highest luminosity at a given temperature. The solid 
points indicate the positions of the VLM dwarfs Gl 406, VB10 and LHS 2924; the
open circles mark the four L-dwarfs with trigonometric parallax measurements.
The uncertainties in luminosity are primarily due to uncertainties in the
bolometric corrections. }

\figcaption{ a: The spectral-type distribution of the 2MASS L-dwarf sample. b: the
temperature distribution predicted for $\Psi(M) \propto M^0$ - 
the vertical lines indicate our estimate of the L0 to L8 temperature r\'egime. c: the
temperature distribution predicted for $\Psi(M) \propto M^{-1}$. d: the
temperature distribution predicted for $\Psi(M) \propto M^{-2}$.
In each case, the solid histogram outlines the total predicted numbers;
the dotted line marks the contribution from objects with M$> 0.07M_\odot$;
the short-dashed line corresponds to a division at 0.055M$_\odot$; and the 
long-dashed line to M=0.03M$_\odot$. This illustrates
the much greater contribution made by low-mass brown dwarfs to the total
if $\Psi(M) \propto M^{-2}$.}

\figcaption {The 2.2 $\mu m$ and bolometric luminosity functions and the effective temperature
distribution predicted by our models with B(t)=constant and $\Psi(M) \propto M^{-\alpha},
\alpha=1,2$. The $\alpha$=1 predictions are plotted as a solid line; the dotted histogram 
plots the expected distribution for $\alpha=2$. The locations of GD 165B and Gl 229B are
indicated as points of reference. All results are scaled to an 8-parsec
radius spherical volume element.}

\setcounter{figure} {0}
\begin{figure}
\plotone{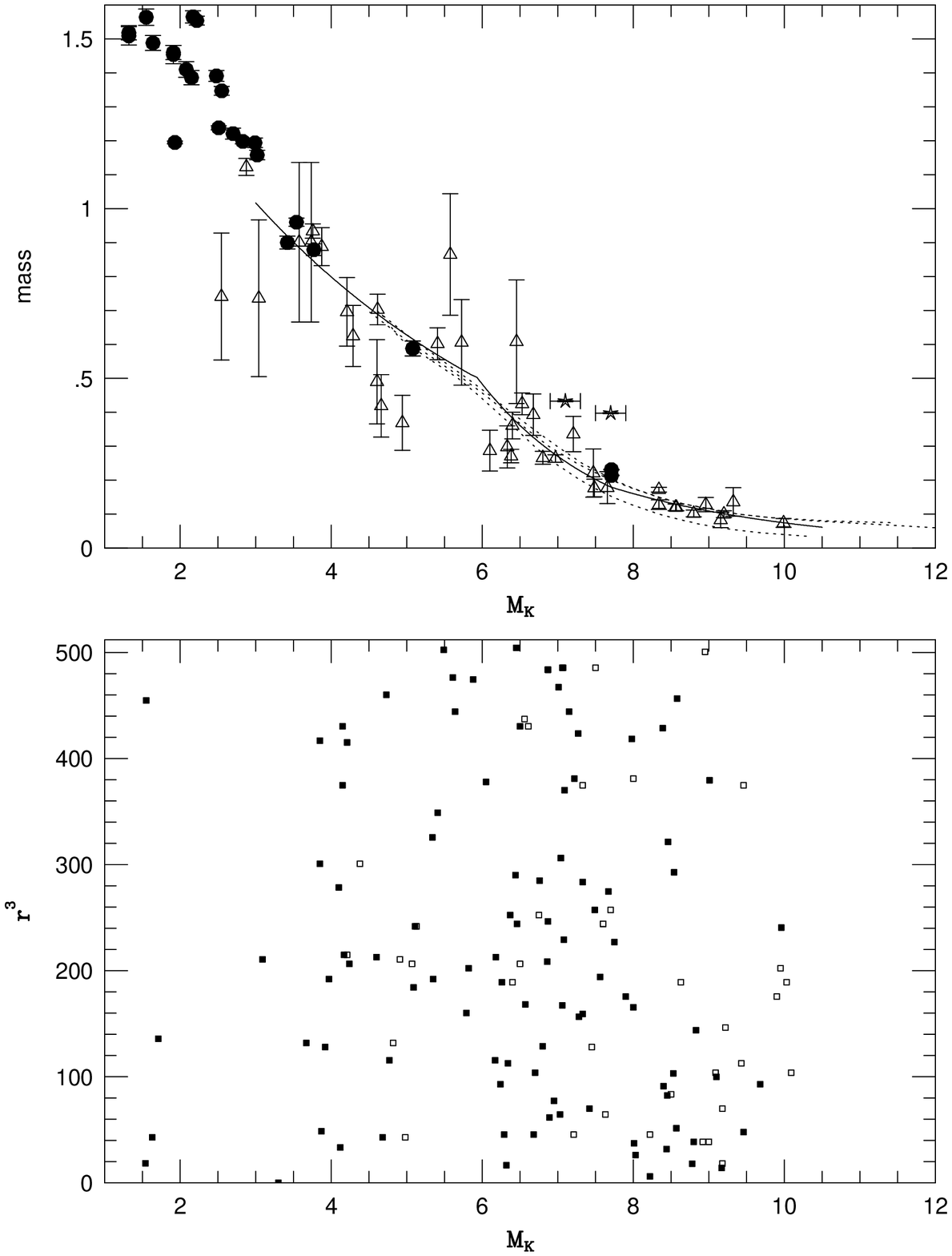}
\caption{}
\end{figure}

\begin{figure}
\plotone{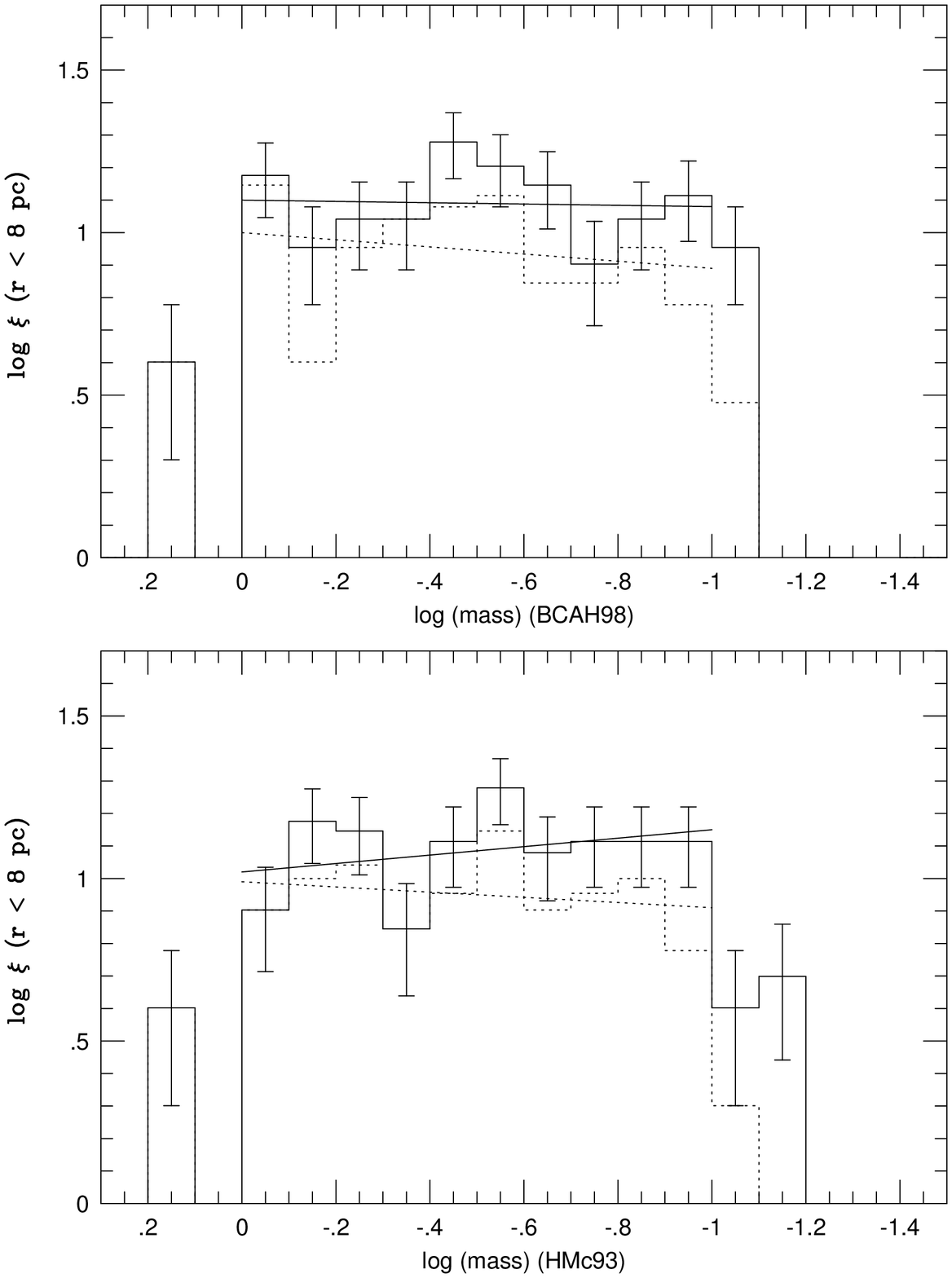}
\caption{}
\end{figure}

\begin{figure}
\plotone{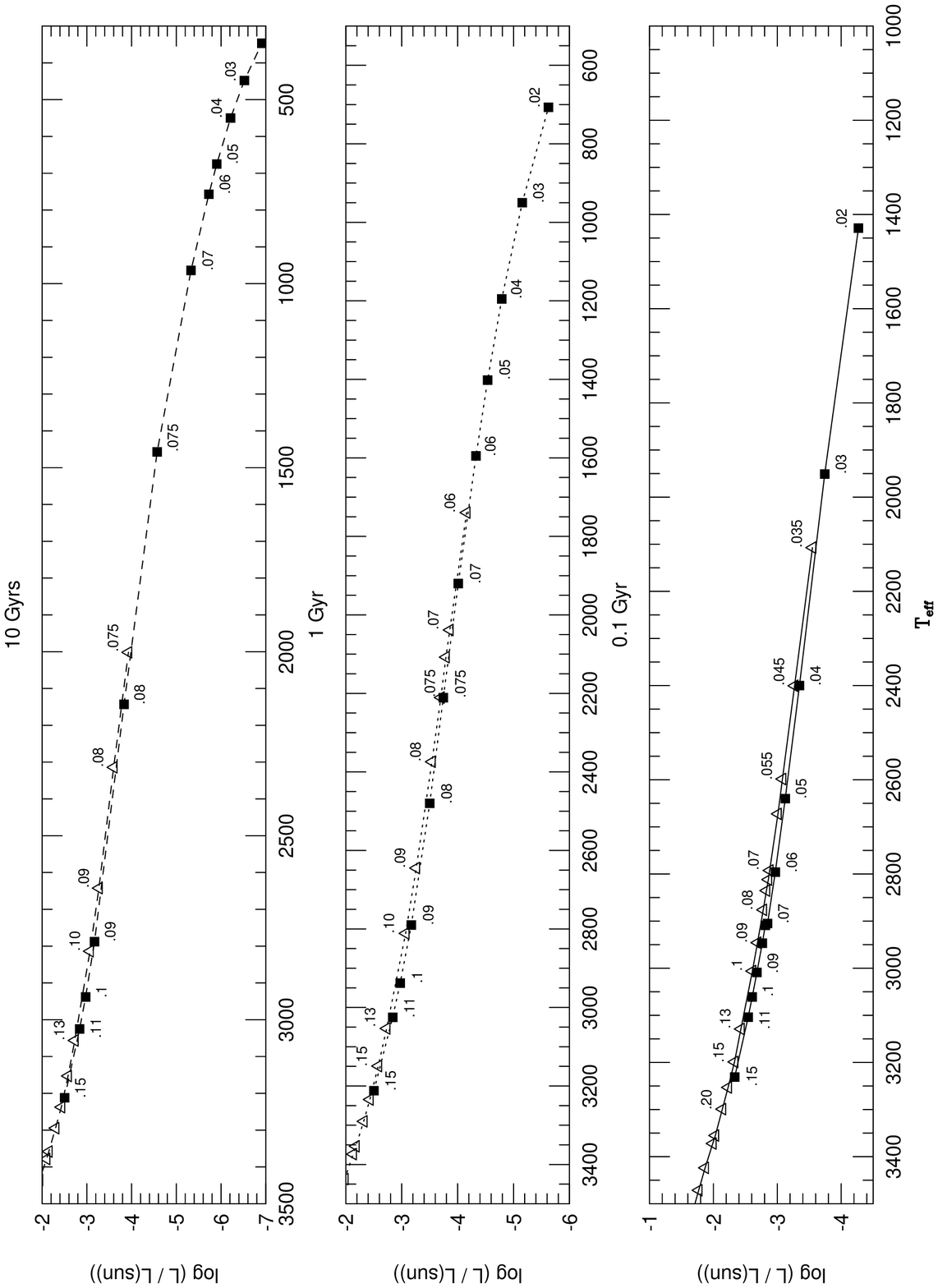}
\caption{}
\end{figure}

\begin{figure}
\plotone{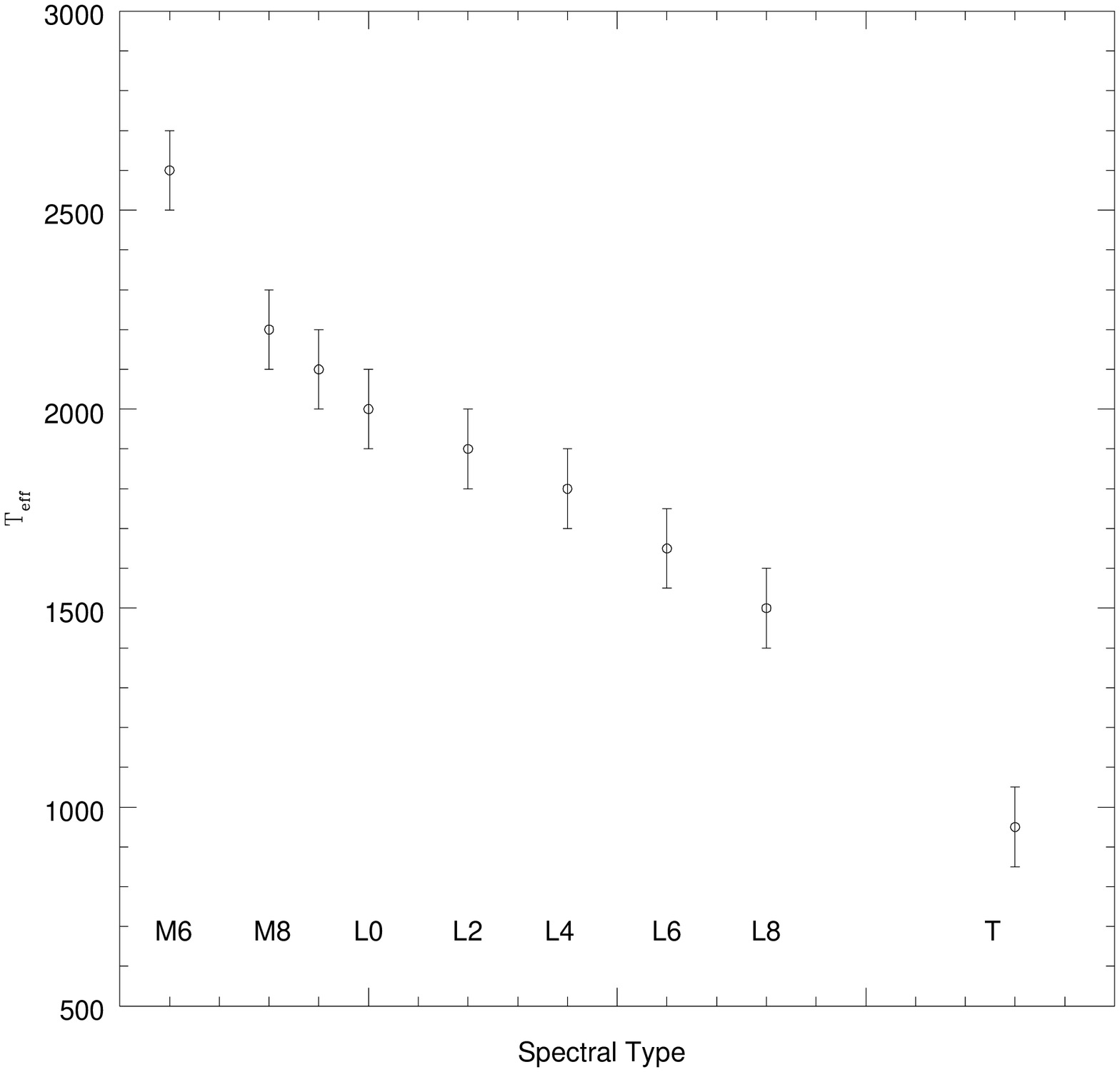}
\caption{}
\end{figure}

\begin{figure}
\plotone{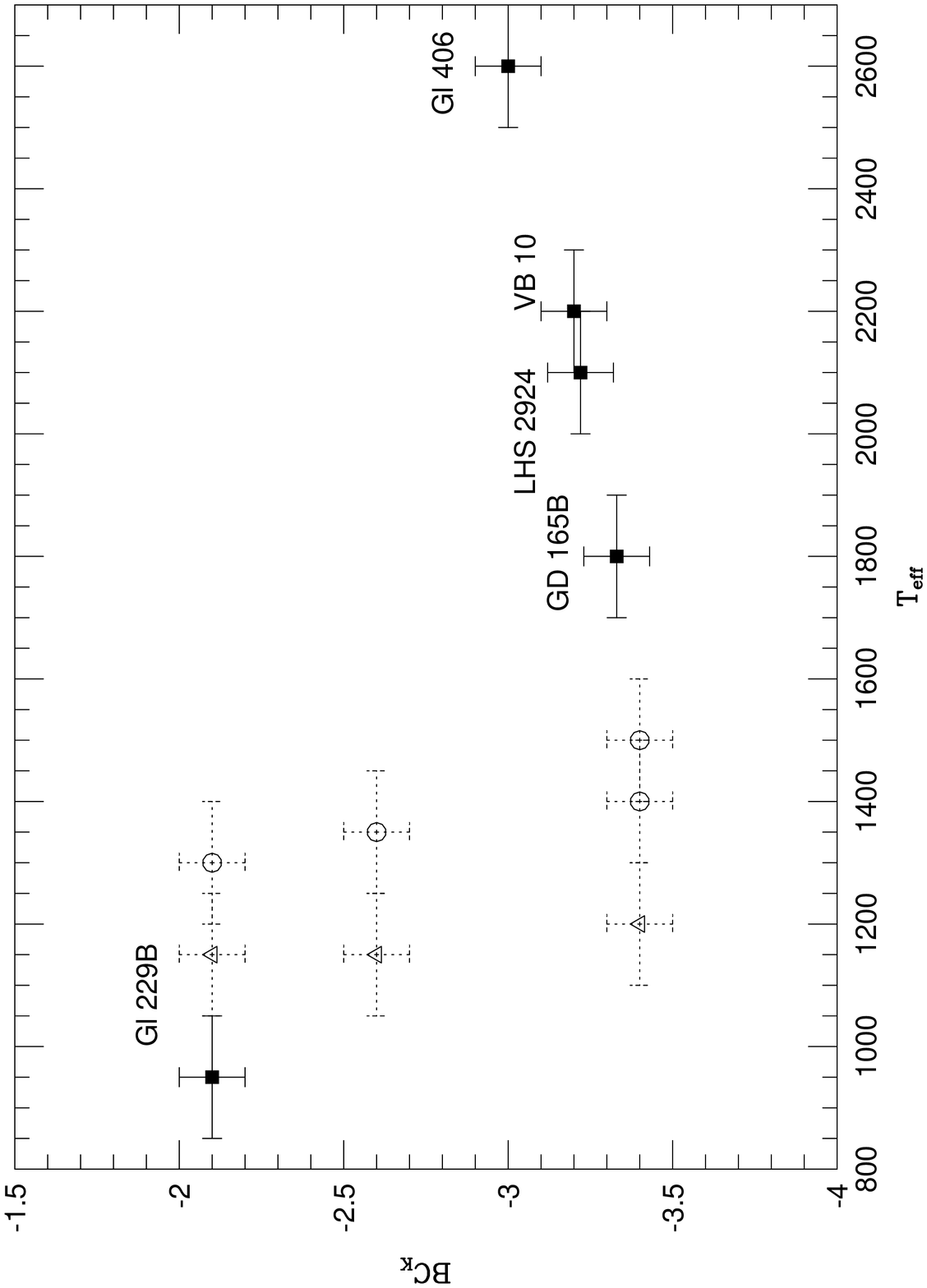}
\caption{}
\end{figure}

\begin{figure}
\plotone{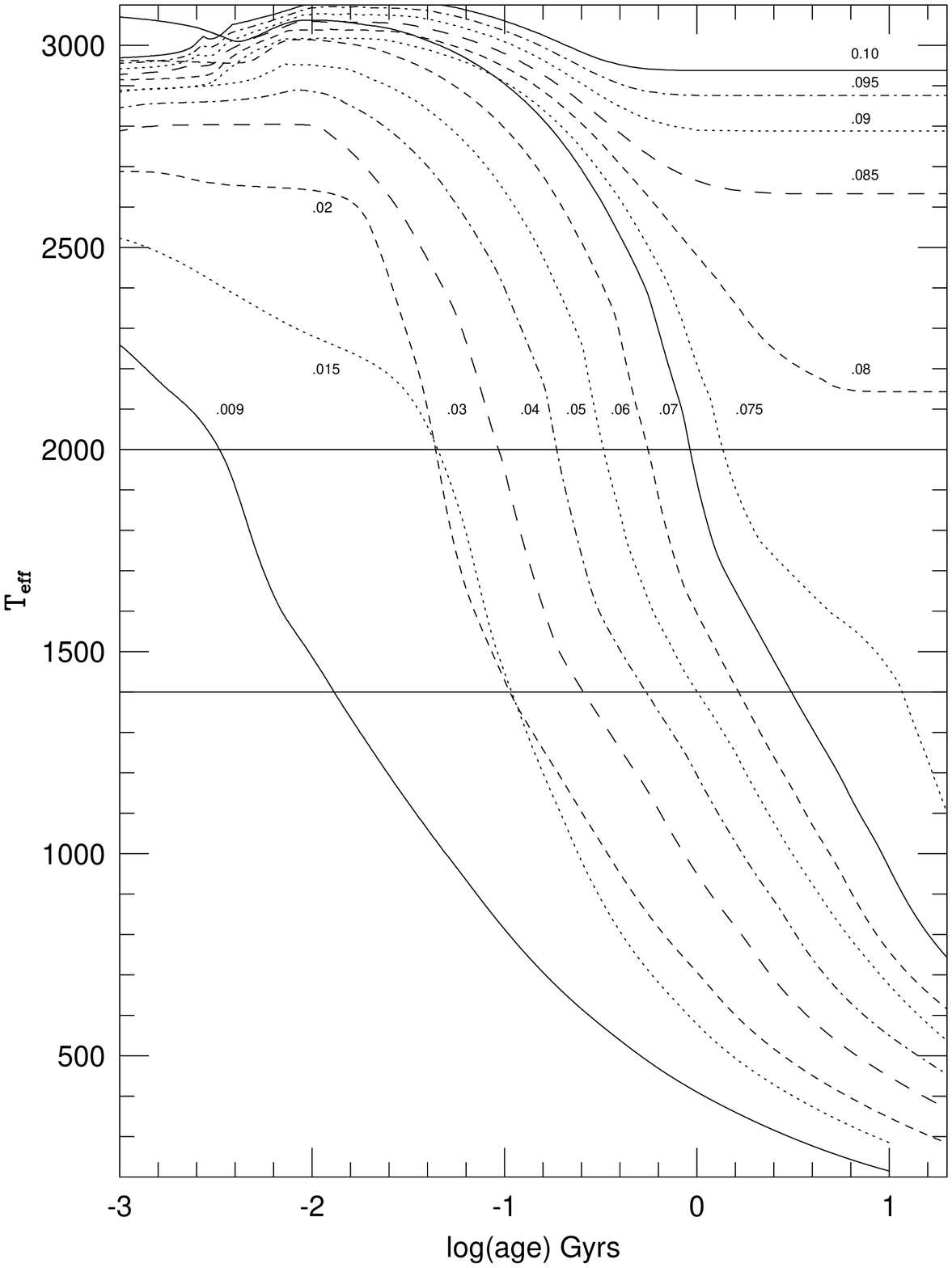}
\caption{}
\end{figure}

\begin{figure}
\plotone{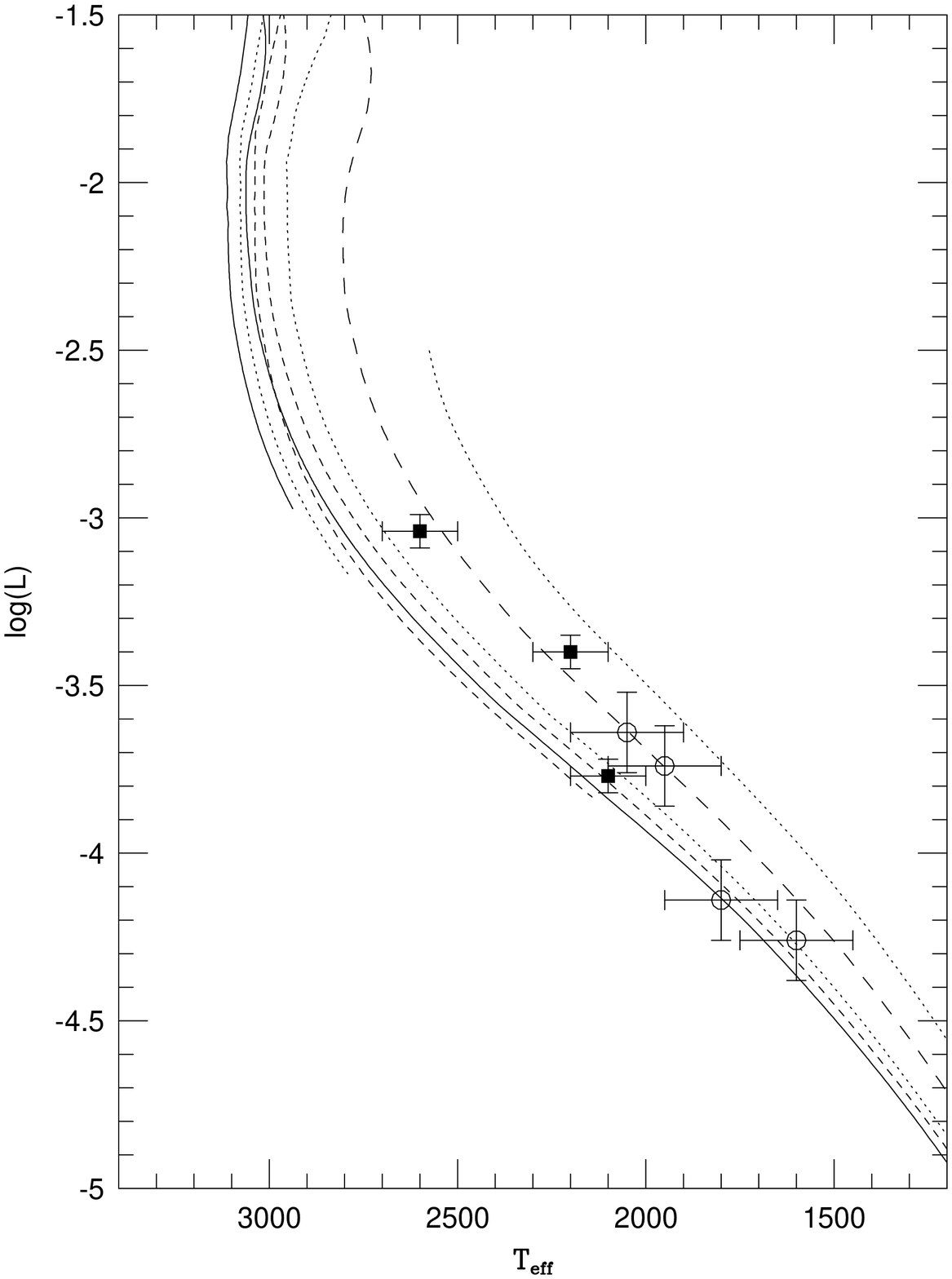}
\caption{}
\end{figure}

\begin{figure}
\plotone{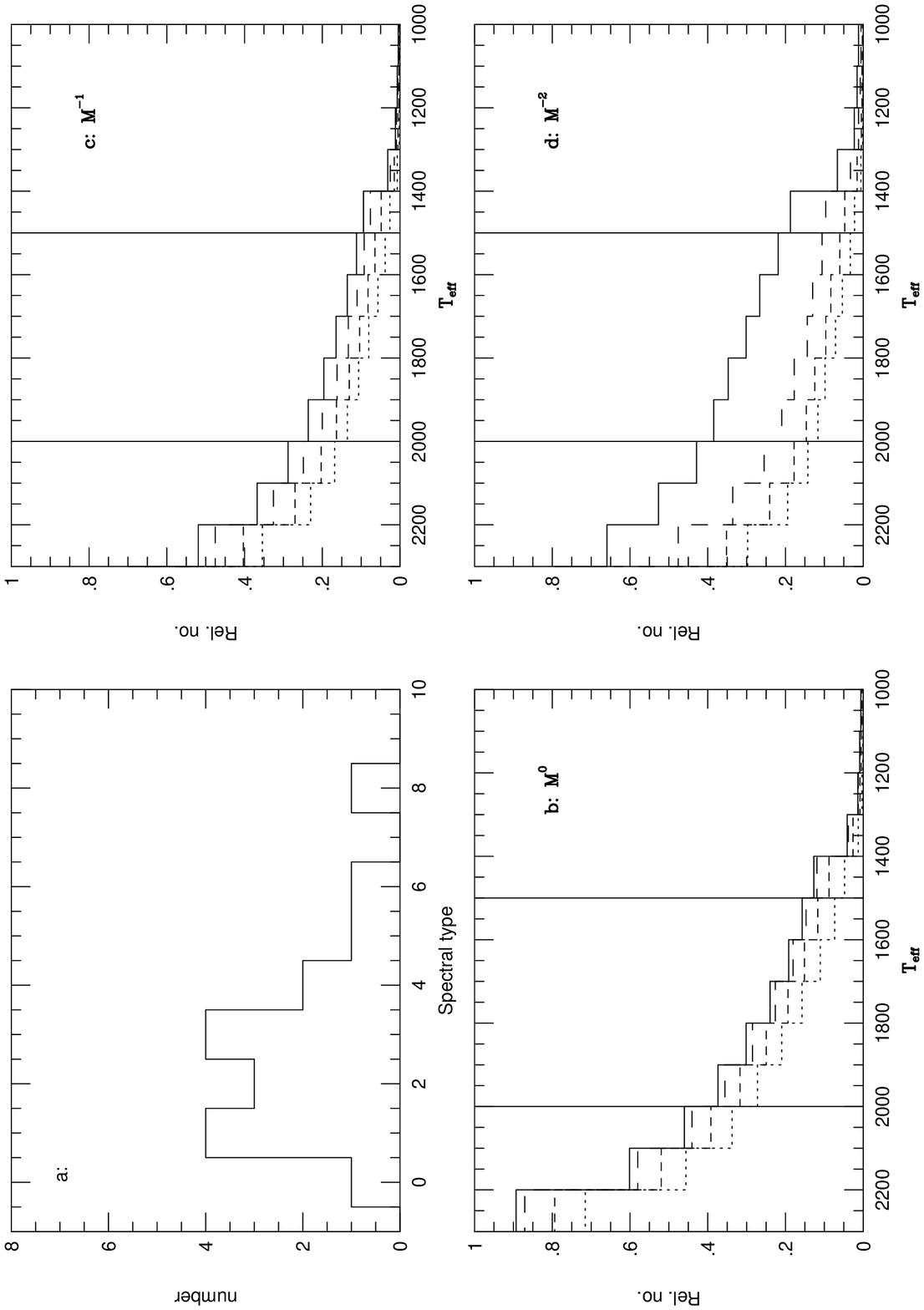}
\caption{}
\end{figure}

\begin{figure}
\plotone{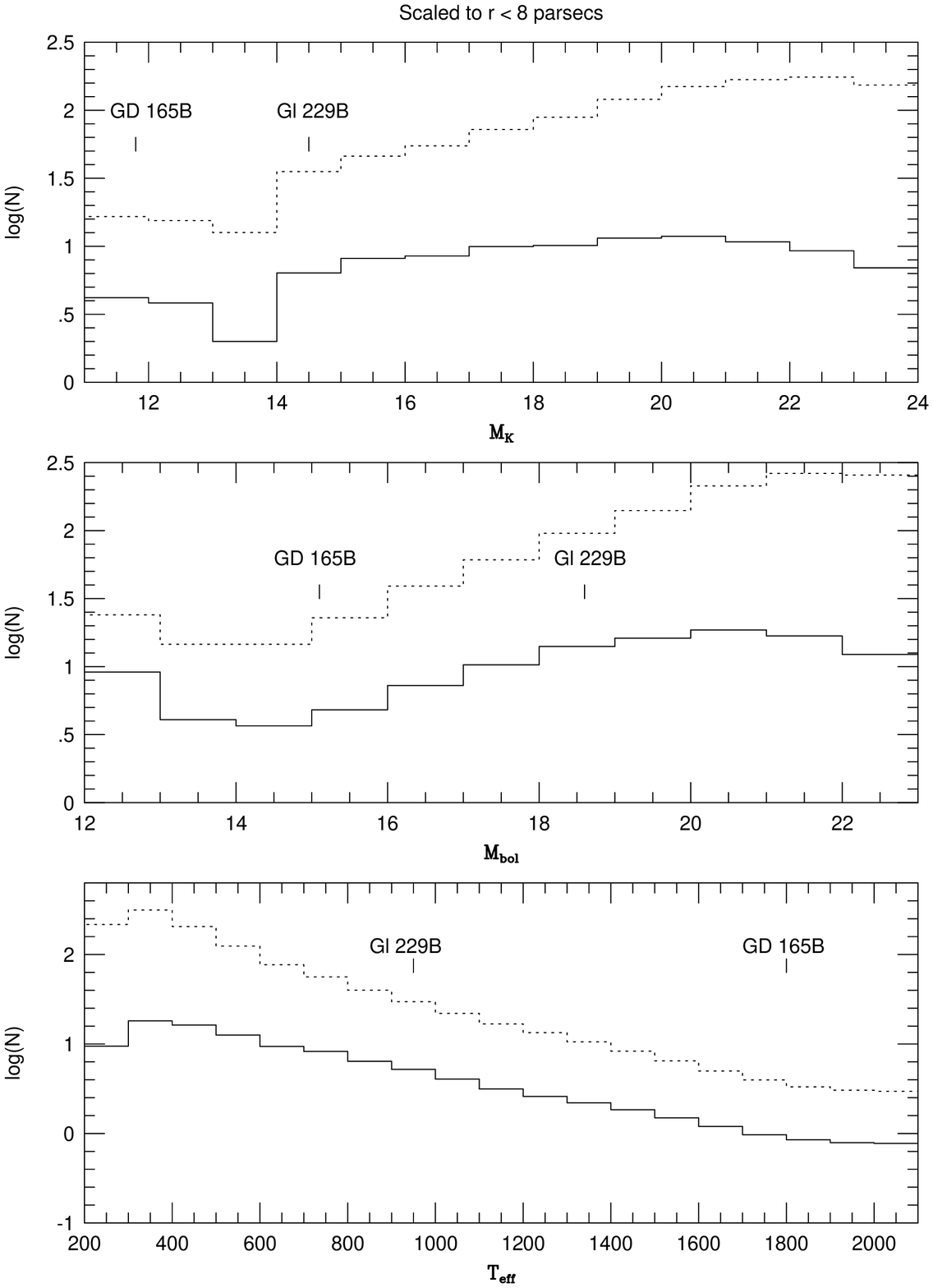}
\caption{}
\end{figure}

\end{document}